\documentclass[aps,prc,preprint,groupedaddress,showpacs,floatfix]{revtex4}

\usepackage{epsfig}
\usepackage{amsmath}

\begin{document}

\preprint{OSU/Yanez-Grazing}

\title{Predicting the production of neutron rich heavy nuclei in multi-nucleon transfer reactions using GRAZING-F}

\author{R. Yanez}
\author{W. Loveland}
\affiliation{Department of Chemistry, Oregon State University, Corvallis, OR, 97331.}
\date{\today}

\begin{abstract}

{\bf Background:}
Multi-nucleon transfer reactions have recently attracted attention as a possible path to the synthesis of new neutron-rich heavy nuclei.

{\bf Purpose:} 
We study transfer reactions involving massive nuclei with the intention of understanding if the semi-classical model GRAZING coupled to an evaporation and fission competition model can satisfactory reproduce experimental data on transfer reactions in which fission plays a role.

{\bf Methods:}
We have taken the computer code GRAZING and have added fission competition to it (GRAZING-F) using our current understanding of $\Gamma_n/\Gamma_f$, fission barriers and level densities.

{\bf Results:}
The code GRAZING-F seems to satisfactory reproduce experimental data for $+1p$, $+2p$ and $+3p$ transfers, but has limitations in reproducing measurements of larger above-target and below-target transfers. Nonetheless, we use GRAZING-F to estimate production rates of neutron-rich $N=126$ nuclei, actinides and transactinides.

{\bf Conclusions:}
The GRAZING code, with appropriate modifications to account for fission decay as well as neutron emission by excited primary fragments, does not predict large cross sections for multi-nucleon transfer reactions leading to neutron-rich \textit{transactinide} nuclei, but predicts opportunities to produce new neutron-rich \textit{actinide} isotopes.

\end{abstract}


\maketitle

\section{Introduction}

Experimentalists have had a long-standing interest in multi-nucleon transfer reactions \cite{schroder84,volkov89} hoping to synthesize new neutron-rich isotopes not normally accessible by neutron capture and fusion reactions \cite{schadel78,schadel82,lee82,moody86,moody87,welch87,gregorich87}. Cross sections of actinides produced in transfer reactions using light and heavy projectiles and actinide targets were measured by the chemical separation of the products in a series of experiments in the late 70's and 80's. The systematic trend that emerged after the series of experiments with U, Cm and Cf targets is that the use of transfer reactions to produce unknown neutron-rich actinides is favorable for below-target species and limited for above-target species. The production of neutron-rich trans-target nuclides up to Fm and Md with cross sections $\sim 0.1 \mu$b were observed. The basic problem in making heavier nuclei was that the higher excitation energies that led to broader isotopic distributions caused the highly excited nuclei to fission.

The interest in transfer reactions has been recently boosted by the prediction of larger than expected cross sections for the production of heavy nuclei, within the framework of a dynamical model based on Langevin equations, by taking advantage of shell effects which may favor a large flow of nucleons resulting in the formation of surviving heavy nuclei \cite{zaggy06,zaggy08}. In this picture, low-energy multi-nucleon transfer reactions of very heavy nuclei, such as U+Cm, may produce one primary reaction product in the vicinity of $Z=82$, $N=126$ closed shells, leaving the second primary product in the actinide or transactinide region with very low excitation energy and thus, with increased probability of surviving fission.  This model was able to account for the previously measured radiochemical data \cite{zagrebaev13}.

The motivation and interest in multi-nucleon transfer reactions in Ref. \cite{zaggy06} and the present paper is two-fold: (a) the possibility of producing the most neutron-rich heavy nuclei for studies using nuclear spectroscopy, atomic physics and chemistry and (b) the difficulty in pursuing the study of nuclei with high atomic numbers using fusion reactions. Traditional ``cold'' fusion reactions have production cross sections of $10-100$ fb beyond $Z=112$, and ``hot'' fusion reactions have cross sections of the order of a few pb for elements $Z\simeq 118$. The upper limit cross sections for $Z=119$ and $Z=120$ have been established to be of the order of 100 fb \cite{chris14}. This difficulty has spurred the renewed interest in low-energy multi-nucleon transfer reactions as a way of accessing new neutron-rich transactinide nuclei that are closer to the ``island'' of stability near the neutron shell $N=184$ not accessible by fusion reactions.

Multi-nucleon transfer reactions in the quasi-elastic and deep-inelastic regimes have been extensively modeled with the semi-classical description of Winther \cite{winther,winther2}, implemented in the computer code GRAZING \cite{grazing}. This code considers the multi-step exchange of nucleons between the colliding nuclei in classical trajectories calculated with a Coulomb plus nuclear interaction. GRAZING is known to have shortcomings, i.e.\ the initial deformations of the nuclei is not taken into account and neutron evaporation from the primary products is the only de-excitation mode considered. As a result, the code has mainly been used to predict yields in light projectile reactions with medium to heavy targets in which the fissility of the reaction products studied is small \cite{dasso94,corradi96,corradi97,corradi99,corradi01,montagnoli02,corradi02a,corradi02,corradi13}. The theoretical formalism of GRAZING is described in depth by Winther \cite{winther,winther2}. An outline of the main ingredients and approximations of the model can be found in the topical review by Corradi, Pollarolo and Szilner \cite{corradi09}. Multi-nucleon transfer reactions have also been studied theoretically using the Fokker-Plank equation \cite{schmidt78}, the finite-range DWBA model \cite{farra96}, the Di-nuclear System model \cite{adamian05}, the time-dependent Hartree-Fock theory \cite{kedziora10}, and the Langevin equations \cite{zaggy06}.

The GRAZING code has recently been informally used to predict yields of products in reactions with planned radioactive beams (EURISOL) and isotope ``factories'' (CARIBU), in some cases with actinide targets. In this paper we present an extension to GRAZING in which not only neutron evaporation from the excited primary products is considered, but also fission competition. With such additions to the code, reactions where fission effectively competes with neutron emission, e.g.\ the U+Cm reaction, can be studied and compared to experimental data and other models.

\section{Neutron evaporation and fission competition}

The competition between neutron emission and fission is simulated with the classical formalism of Vandenbosch and Huizenga \cite{vandenbosch},

\begin{equation}
\label{eq1}
\frac{\Gamma_n}{\Gamma_f}=\frac{4A^{2/3}a_f(E^*-B_n)}{K_0a_n(2\sqrt{a_f(E^*-B_f)}-1)}\exp{\left(2\sqrt{a_n(E^*-B_n)}-2\sqrt{a_f(E^*-B_f)}\right)}
\end{equation}

\noindent
where $a_n$ and $a_f$ are the level density parameters at the equilibrium deformation and saddle point, respectively, $B_n$ is the neutron separation energy, $B_f$ is the fission barrier and $K_0=\hbar^2/2mr_0^2$. The fission barriers $B_f$ are taken to be the sum of the Thomas-Fermi barrier \cite{myers99} plus the shell correction term,

\begin{equation}
B_f = B_f^{LD} + U_{shell}
\end{equation}

\noindent
$U_{shell}$ is taken to be the microscopic energy of the Finite Range Droplet Model (FRDM) \cite{moller95}. Angular momentum $J$ is treated by reducing the available energy in Eq.~\ref{eq1} by the rotational energy $E_r$ of the fissioning nucleus and scaling the Thomas-Fermi fission barrier with the Sierk barrier \cite{sierk86}.

The fade-out of the shell correction with increasing excitation energy is treated through the level density parameter following the method of Ignatyuk et al. \cite{ignatyuk75},

\begin{equation}
a(U) = \tilde{a}\left(1+f(U)\delta W /U\right)
\end{equation}

\noindent
where $U$ is the excitation energy, $\delta W=M_{exp}(Z,A)-M_{LD}(Z,A,\alpha)$ is the difference between the experimental mass and the theoretical mass within the FRDM (the shell correction to the mass formula), and,

\begin{equation}
f(U)=1-\exp(-\lambda U)
\end{equation}

\noindent
is a semi-empirical formula that drives the energy dependence of $a$. The asymptotic level density parameter $\tilde{a}$ is given by,

\begin{equation}
\tilde{a}=\alpha A + \beta A^{2/3} \tilde{s}
\end{equation}

\noindent
where $\tilde{s}$ is the surface on the nucleus in units of the equivalent-size sphere. The nuclear surface area $S$ is estimated using the standard expansion of the nuclear radius in spherical harmonics, which for symmetric deformations (as in a nucleus undergoing fission) and ignoring higher order terms, is given by,

 \begin{equation}
S=4\pi R_0^2 \left [ 1 + \frac{2}{5} a_2^2 - \frac{4}{105} a_2^3 + \dots \right ]
\end{equation}

\noindent
where

\begin{equation}
a_2 = \left( \frac{5}{4\pi} \right)^{1/2} \beta_2
\end{equation}

\noindent
and $\beta_2$ is the calculated quadrupole deformation of the nuclear ground state within the FRDM\@. We use the coefficients obtained with a realistic Wood-Saxon potential \cite{ignatyuk75},

\begin{eqnarray*}
\alpha = 0.073, \beta = 0.095, \gamma = 0.061 &\text{MeV}^{-1}
\end{eqnarray*}

\begin{figure}[hbp]
\includegraphics[scale=0.7]{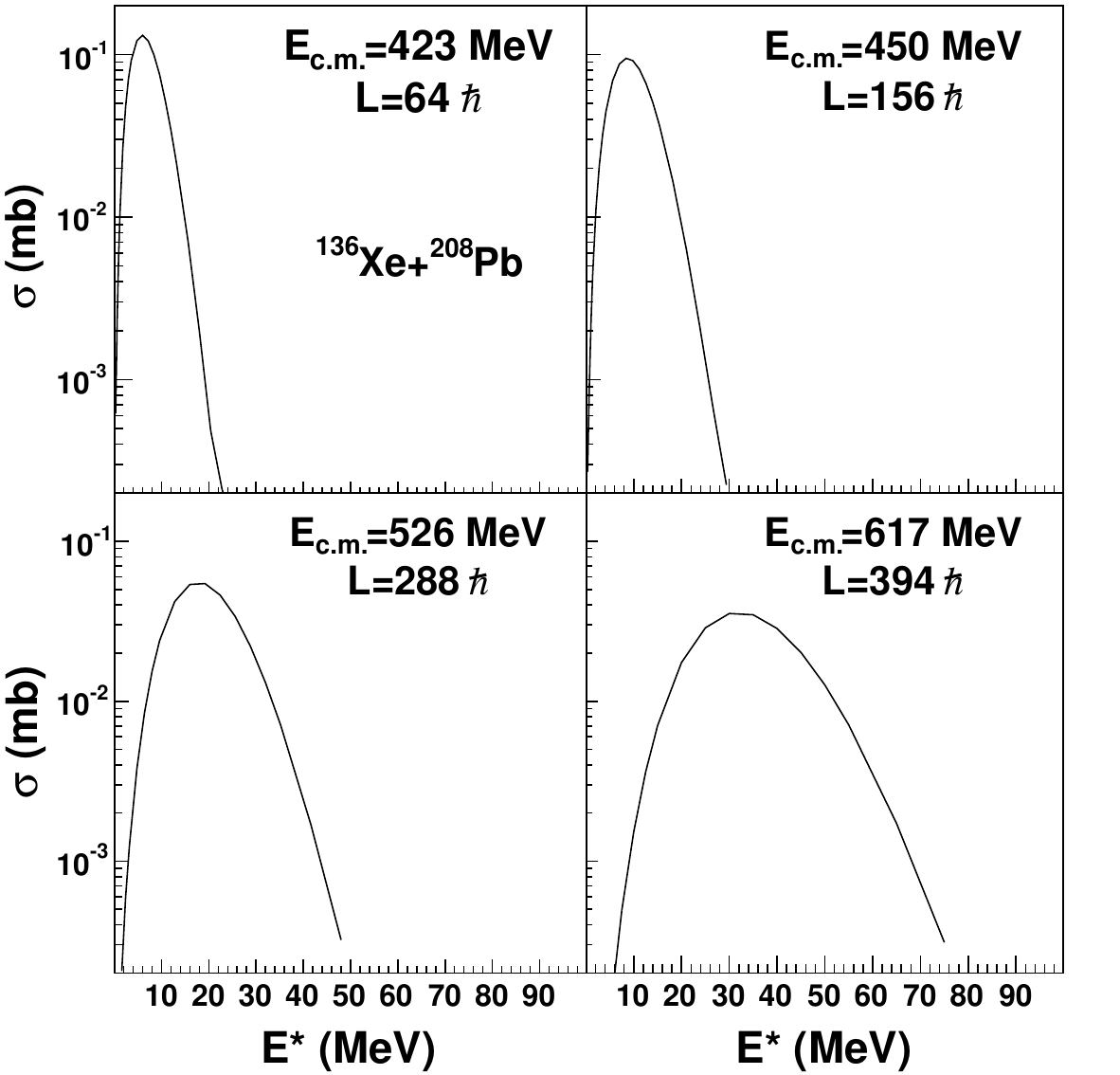}
\caption{Excitation energy distributions predicted by GRAZING in the reaction $^{136}$Xe+$^{208}$Pb at $E_{c.m.}=423, 450, 526$ and $617$ MeV for the partial wave leading to the highest cross section for producing primary product $^{204}_{78}$Pt$_{126}$. The partial wave $L$ is given in the panels.}
\label{fig_exciprob}
\end{figure}

The present simulations take the output of GRAZING in the form of the excitation energy $E^*$ distributions of primary products ($Z$,$A$) for each partial wave, which is converted into a discrete cumulative probability function, which in turn is used to numerically select an event with the generation of a single random number. Fig.~\ref{fig_exciprob} shows the simulated excitation energy distribution in the $^{136}$Xe+$^{208}$Pb reaction for the partial wave leading to the highest cross section for producing primary product $^{204}_{78}$Pt ($N=126$) at $E_{c.m.}=423, 450, 526$ and $617$ MeV. The most probable $E^*$ is $6.0, 8.4, 19.2$ and $30.0$ MeV, for partial waves $L=64, 156, 288$ and $394 \hbar$, respectively. The average transferred angular momentum $J$ at the most probable $E^*$ is $0.031, 0.23, 1.1$ and $3.5 \hbar$, respectively.

For each initial event ($Z$,$A$,$E^*$,$J$), ${\Gamma_n}/{\Gamma_f}$ is calculated using Eq.~\ref{eq1} assuming $a_f=a_n$. The calculated ${\Gamma_n}/{\Gamma_f}$ is tested with a random number to decide whether neutron evaporation or fission happens. If fission happens, the testing of the event is terminated. If neutron evaporation happens, $A$ is decreased by one mass number and $E^*$ is reduced by ($B_n+E_n$), where $B_n$ is the neutron binding energy and $E_n$ is the neutron kinetic energy sampled randomly with a Maxwellian probability function of nuclear temperature $T=\sqrt{aE^*}$,

\begin{equation}
E_n=-T*\left[ \log(r_1)+\log(r_2) \right]
\end{equation}
\noindent
where $r_1$ and $r_2$ are two independent random numbers. If $J>0$, it is assumed the evaporated neutron carries $1\hbar$ of angular momentum. The procedure is iterated until $E^*<B_n$.

Each simulation is performed with the standard set of parameters of GRAZING \cite{grazing} and the de-excitation part is simulated with $10^{12}$ cascades in a High-Performance Computing Cluster using 40 nodes. This large number of cascades is necessary in order to simulate events with the lowest cross sections. The angular momentum transferred to the primary products is rather modest and it is therefore assumed that $J=0 \hbar$ in all simulations except where otherwise indicated.

In what follows we refer to the simulations described in this section as GRAZING-F. 

\section{Comparison with experimental data}

We have gathered an extensive set of experimental data to compare with simulations. The reactions we have studied can be divided into two categories; reactions in which the target is a Pb-like nucleus or is an actinide. In the former case, the fissility of the primary fragments is relatively low and fission may be relevant only in the case of very high excitation energy.

Some of the experimental studies have been done with very thick targets, which pose a difficulty when comparing to simulation since the reported cross section represents an integrated quantity between the incident and exit projectile energies. If the projectile stops in the target material, the cross section represents an integrated quantity down to the interaction barrier of the reaction. For thin-target experiments (for which the projectile exits the target), the projectile energy used in the simulations was assumed to be the effective mid-target projectile energy, estimated with range tables \cite{northcliffe}. For thick-target experiments (for which the projectile stops in the target), the simulations were done in suitable slices of the effective target thickness (the range up to the interaction barrier) and the cross section was calculated as the weighted mean of the slice cross section simulated at the mid-slice energy.

We have studied only the yields of surviving target-like products. Table~\ref{tab1} lists the reactions simulated, the interaction barriers, the simulated transfer and transfer-fission cross sections. The last column of the table is the reference to the experimental data used in the comparisons.

\subsection{The $^{238}$U+$^{238}$U,$^{248}$Cm reactions}

The $^{238}$U+$^{238}$U and $^{238}$U+$^{248}$Cm reactions were studied in the late 70s (Ref. \cite{schadel78}) and 1982 (Ref. \cite{schadel82}), to determine the feasibility of using multi-nucleon transfer and deep-inelastic reactions to synthesize superheavy elements. $^{238}$U beams bombarded thick $^{238}$U and $^{248}$Cm targets and radiochemical methods were employed to deduce cross sections of actinide isotopes. The experimental data was later reexamined by Kratz \textit{et al.} \cite{kratz13}. The data reported in this latter paper form the basis of the present comparison with simulations. The two systems have also been modeled within the diffusion model \cite{reidel79} and the dynamical model based on the multi-dimensional Langevin equations \cite{zaggy06}. 

\begin{figure}[hbp]
\includegraphics[scale=0.5]{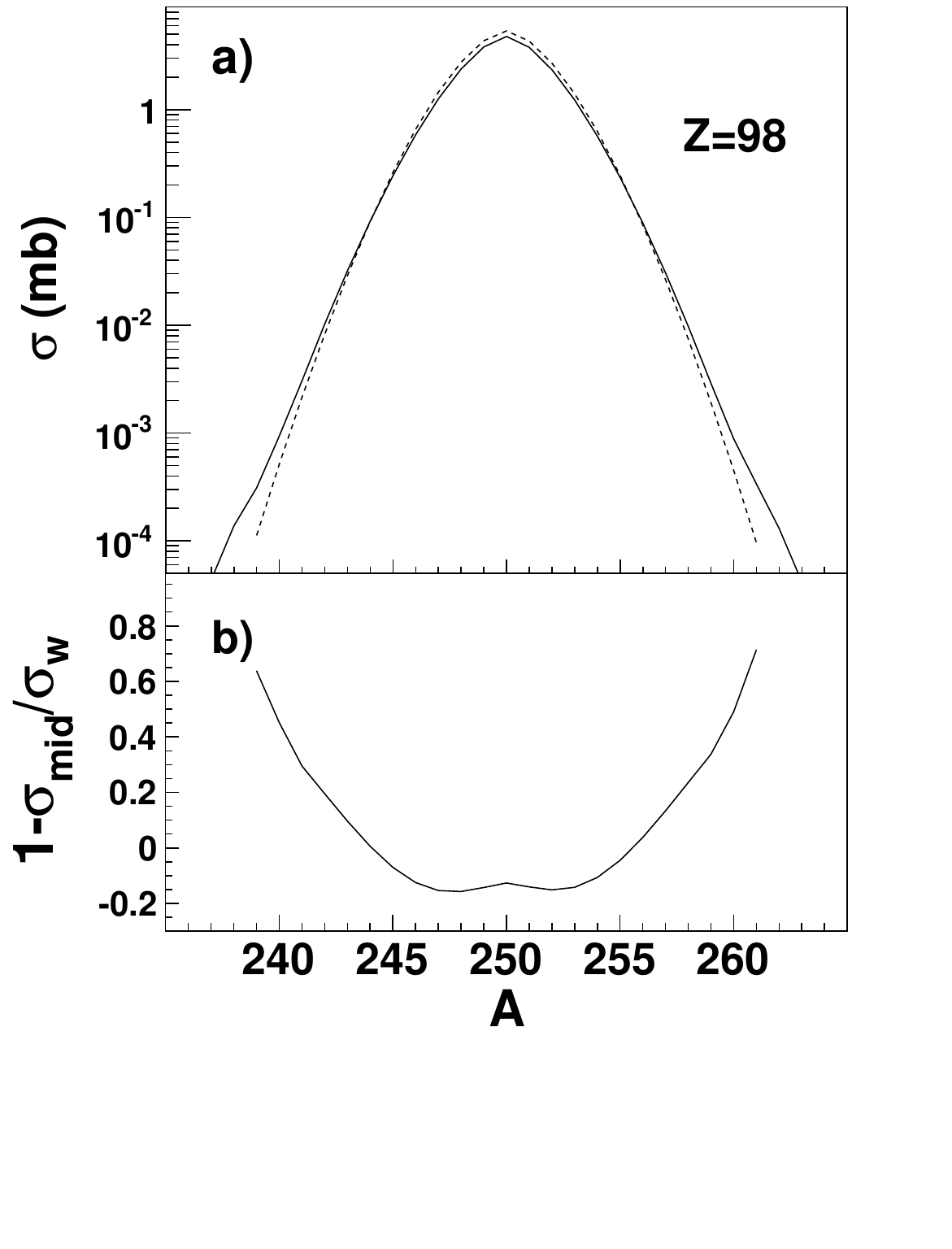}
\caption{Panel a) shows a comparison between simulations of the primary yields of $Z=98$ products in the $^{238}$U+$^{248}$Cm reaction with GRAZING by constructing a weighted cross-section (solid line) and a single simulation at the mid-target energy (dashed line.) The effective target thickness is $4.8$ mg/cm$^2$ and the weighted simulation is made by assuming a stack of ten identical target slices. Panel b) shows the deviation.}
\label{fig_ratio}
\end{figure}

The $^{238}$U+$^{248}$Cm reaction was experimentally studied at entering projectile energy of $1760$ MeV with a target thick enough to stop the projectiles. The mid-target energy is $1650$ MeV. For the purpose of the simulations, the effective target thickness ($4.8$ mg/cm$^2$) was divided in ten equal slices (equivalent to a stack of ten thin targets of $0.48$ mg/cm$^2$ each) and GRAZING was run at the equivalent mid-slice energy. The upper panel of Fig.~\ref{fig_ratio} shows a comparison between the cross section obtained by weighting the ten yields of $Z=98$ primary products (solid line) and the yield resulting from the effective mid-target energy alone (dashed line.) The weighted distribution is broader because it includes partial distributions at higher energies. The lower panel in Fig.~\ref{fig_ratio} shows the deviation. Assuming a single mid-target energy for this reaction may result in a systematical error of $\sim 10$\% around the most probable mass, and more than 50\% at the extremes. This result justifies the use of the weighted procedure at the expense of considerable computing time.

\begin{figure}[hbp]
\includegraphics[scale=1.2]{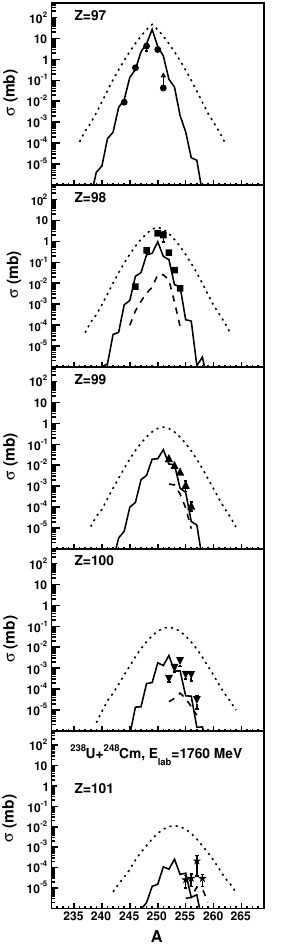}
\caption{Cross sections of surviving nuclei in the reaction $^{238}$U+$^{248}$Cm at entering projectile energy $E_{lab}=1760$ MeV. Experimental data from Ref.~\cite{kratz13} are shown as solid symbols and the predicted cross sections (GRAZING-F) as solid lines. The measurement of $^{251}$Bk is a lower limit which is indicated with an arrow. Dotted lines show the predicted yields of primary products. Dashed lines shows the Langevin simulations in Ref.~\cite{zaggy06}}
\label{fig_schadel1}
\end{figure}

Fig.~\ref{fig_schadel1} shows the direct comparison between experimental data and simulations with GRAZING-F for actinides with $Z$=97-101 in the $^{238}$U+$^{248}$Cm reaction. The experimental data of Ref. \cite{schadel78} is plotted as solid symbols, the simulated yield of surviving products as solid lines and the primary product yields as dotted lines. The experimental cross section for $^{251}$Bk is a lower limit, which is denoted by upward arrow in the panel for $Z=97$. The agreement for $+1p$, $+2p$ and $+3p$ transfers is remarkable, whereas the simulation is less successful for larger $p$ transfers. The reason for this discrepancy is that GRAZING predicts insufficient primary neutron transfers for these nuclei, as can be seen by comparing the simulated primary and secondary yields. The larger excitation energy moves the surviving product distribution towards lower $A$ values. For the simulation to reproduce the yields of Fm and Md nuclei at least 2 and 4 additional neutrons would, on average, be required to be transferred to the primary products. The odd-even effects displayed by the simulations are a direct consequence of the odd-even effects introduced in the calculation of ${\Gamma_n}/{\Gamma_f}$ in the de-excitation stage. The yields predicted by the Langevin model \cite{zaggy06} are shown as dashed lines in Fig.~\ref{fig_schadel1} (see Fig. 4 in Ref.~\cite{zaggy06}.)

\begin{figure}[hbp]
\includegraphics[scale=0.5]{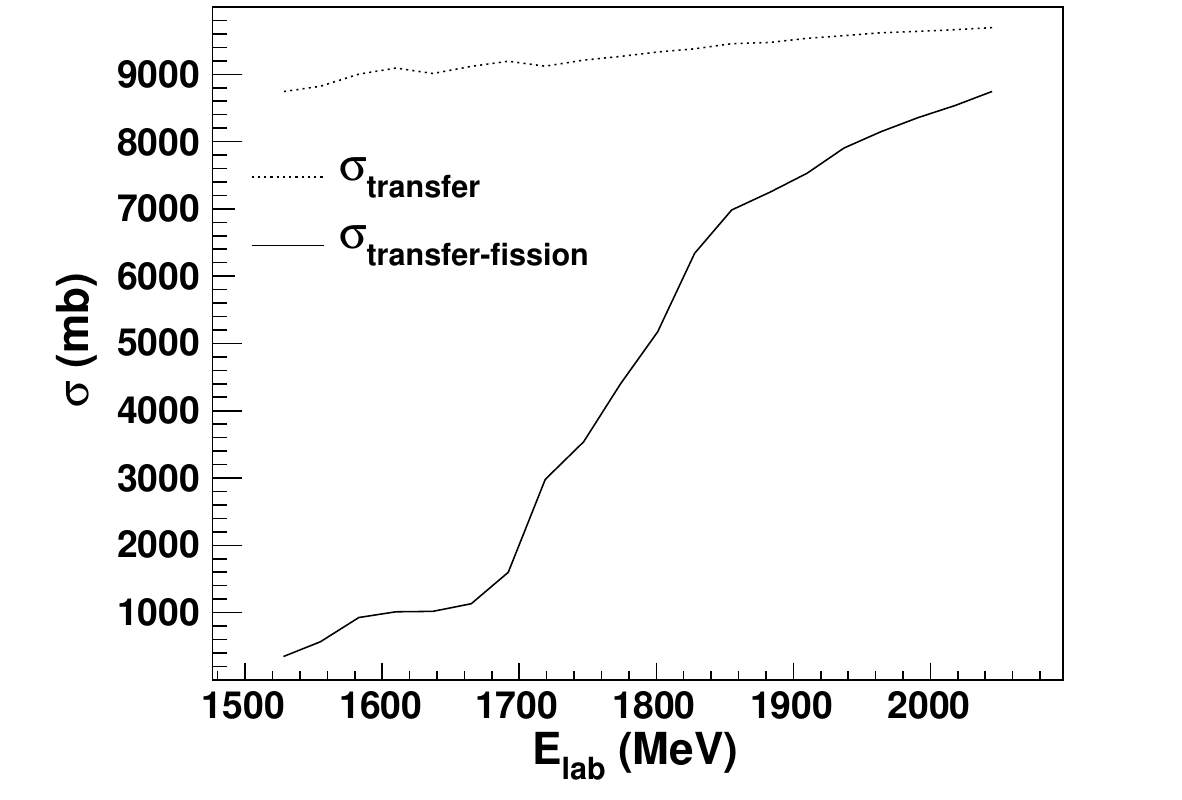}
\caption{Predicted transfer (GRAZING) and transfer-fission cross sections (GRAZING-F) as a function of mid-slice energy in the $^{238}$U+$^{238}$U reaction at entering energy $E_{lab}=2059$ MeV.}
\label{fig_u_u_fission}
\end{figure}

The $^{238}$U+$^{238}$U reaction was experimentally studied at four energies for which independent yields are reported. The targets used were thick enough to stop the projectiles. The entering energies were $2059$ ($11.6$), $1785$ ($5.7$), $1628$ ($2.4$) and $1545$ ($0.65$) MeV (mg/cm$^2$), with mid-target energies of $1787$, $1650$, $1571$ and $1530$ MeV, respectively. Given in parenthesis is the effective target thickness. For the purpose of simulations with GRAZING-F, the effective target thicknesses were divided in 20, 10, 5 and 1 slice(s), respectively. In the $2059$ MeV reaction, the transfer-fission cross section in the first slice is $\sim 90$\% of the transfer cross section and decreases to $\sim 4$\% in the last slice. Fig.~\ref{fig_u_u_fission} shows the transfer and transfer-fission cross section as a function of mid-slice projectile energy. The dependency of the transfer-fission cross section on energy is determined by both the excitation energy distribution of the primary fragments and $\Gamma_{n}/\Gamma_{f}$.

\begin{figure}[hbp]
\includegraphics[scale=1.0,angle=90]{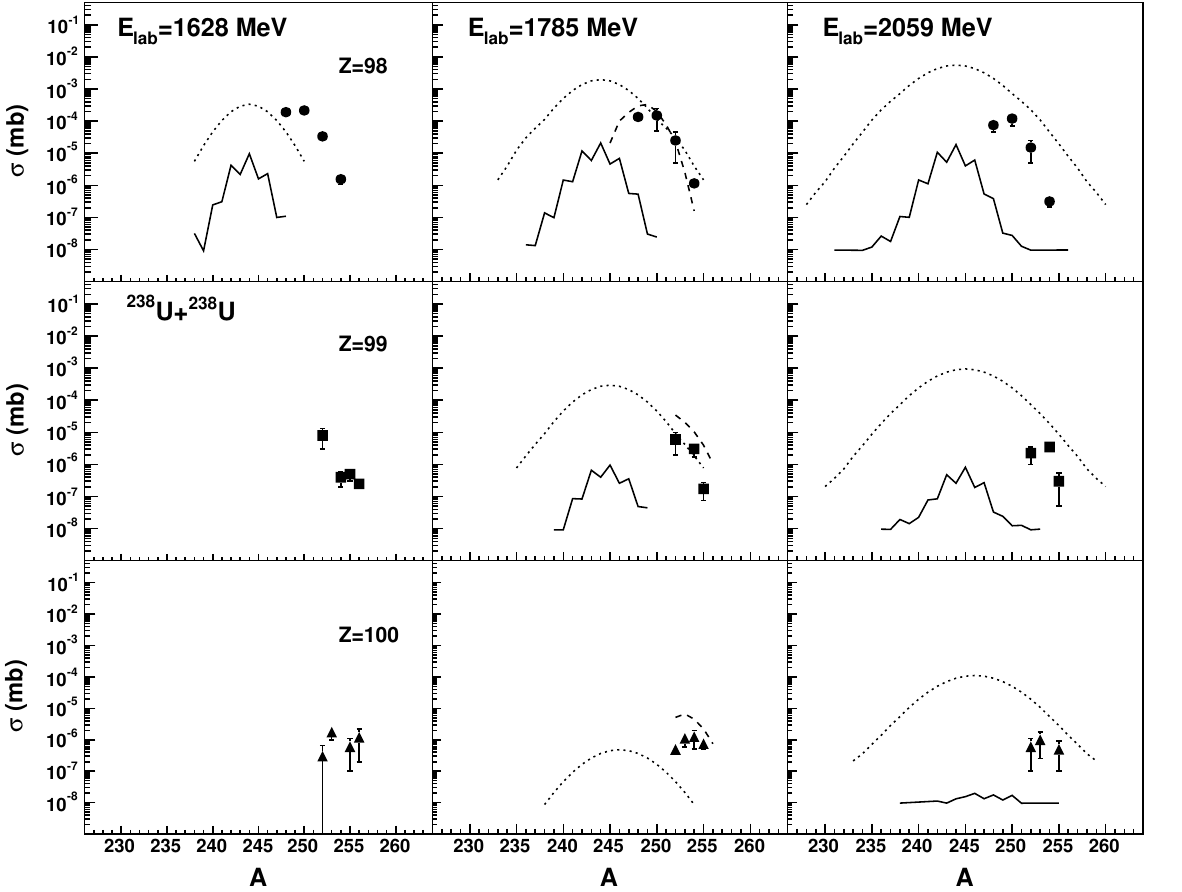}
\caption{Cross sections of surviving nuclei in the reaction $^{238}$U+$^{238}$U at $E_{lab}=1628, 1785$ and $2059$ MeV. Experimental data from Ref.~\cite{kratz13} are shown as solid symbols and predicted cross sections (GRAZING-F) as solid lines. Dotted lines show the predicted yield of primary products.  Dashed lines shows the Langevin simulations in Ref.~\cite{zaggy06}.}
\label{fig_schadel2}
\end{figure}

Fig.~\ref{fig_schadel2} shows the results for the $^{238}$U+$^{238}$U reaction. The simulations do not reproduce the data. GRAZING does not predict $> +5p$ transfers at the lowest energy (not shown in Fig.~\ref{fig_schadel2} for that reason), and $> +6p$ transfers in the $1628$ MeV reaction. As in the case of the $^{238}$U+$^{248}$Cm reaction, GRAZING seems to underpredict the flow of neutrons in $> +4p$ transfers. In the $^{238}$U+$^{238}$U reaction, at least 5 additional neutrons would on average be required to be transferred to the primary products in order for GRAZING-F to reproduce the locations of the maximum yields. The yields predicted by the Langevin model \cite{zaggy06} for the $E_{lab}=1785$ MeV reaction are shown as dashed lines in Fig.~\ref{fig_schadel2}. Comparing both simulations in the two reactions, we may conclude that the Langevin model seem to reproduce fairly well the yields of massive transfers ($> +5p$), as in the $^{238}$U+$^{238}$U reaction, whereas GRAZING-F seems to better reproduce the yields of a few-nucleon transfers ($< +4p$), as in the $^{238}$U+$^{248}$Cm reaction (see Fig.~\ref{fig_schadel1}.)

\begin{figure}[hbp]
\includegraphics[scale=0.5]{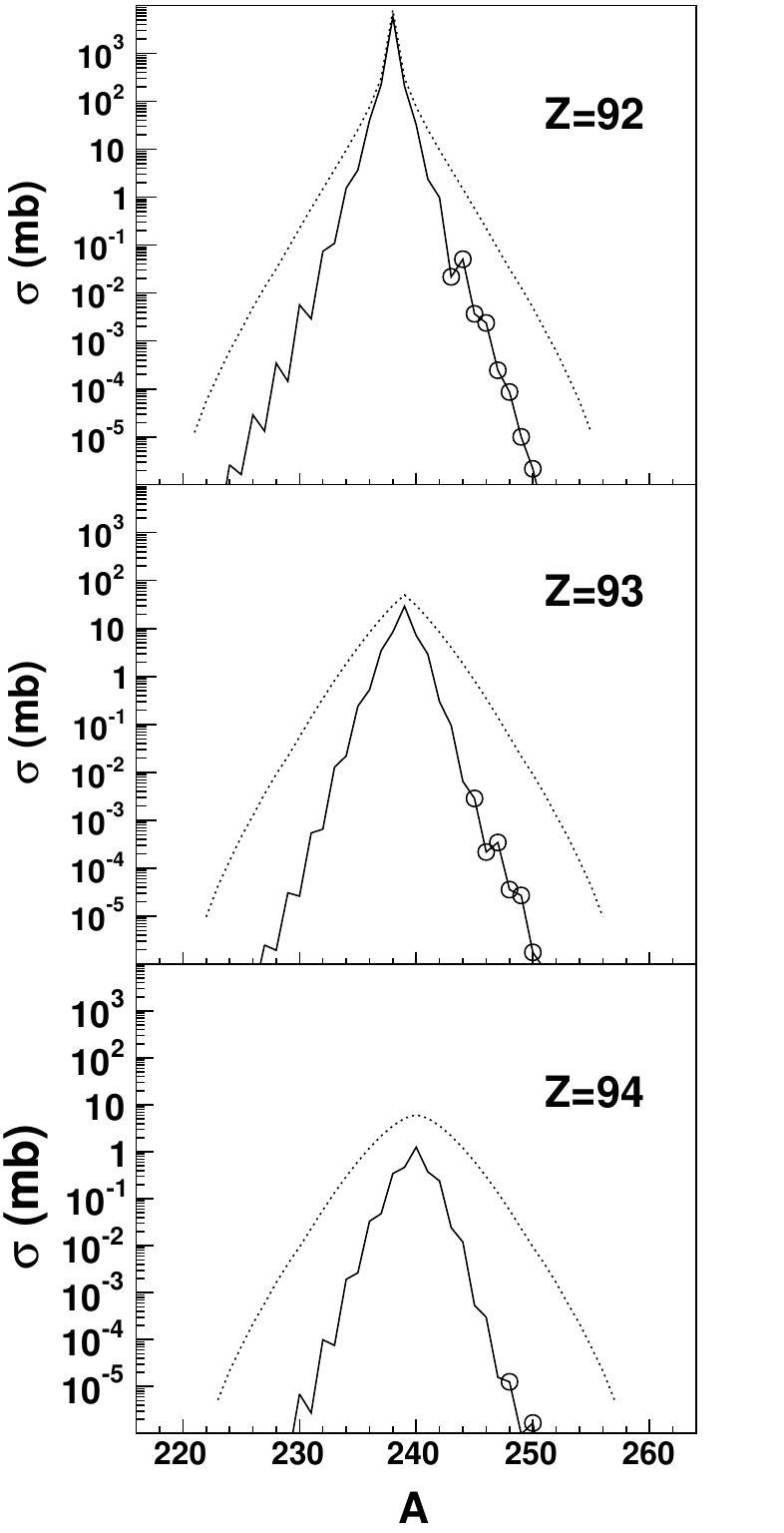}
\caption{Predicted (GRAZING-F) cross sections of surviving nuclei with $Z=92-94$ in the $^{238}$U+$^{238}$U reaction at entering energy $E_{lab}=2059$ MeV. Unknown isotopes are shown as open circles.}
\label{fig_u_u_actinide_yields}
\end{figure}

If we assume GRAZING-F is able to reproduce the yields of $< +4p$ transfers reasonably well, then GRAZING-F predicts substantial yields of unknown actinides in the $^{238}$U+$^{238}$U reaction at $E_{lab}=2059$ MeV. Fig.~\ref{fig_u_u_actinide_yields} shows the predicted production cross sections of $Z=93-94$ isotopes. Open circles represent unknown actinides. The predicted cross sections that are measurable ($> 100$ nb) are listed in Table~\ref{tab2}.

\subsection{The $^{129,132,136}$Xe+$^{248}$Cm reaction}

The $^{129,132}$Xe+$^{248}$Cm reactions were measured in order to study the influence of the projectile $N/Z$ ratio in the production of actinides \cite{welch87}. This work used a thin Cm target and the simulations were therefore done at the mid-target energy of the projectile. The cross sections in the $^{136}$Xe+$^{248}$Cm reaction were measured by chemical separation with the intent to determine the formation cross sections of unknown actinides \cite{moody86}. In this case the simulation was done at energy $E_{lab}=769$, which we have estimated to be the effective mid-target energy for the reaction with entering energy of 790 MeV.

\begin{figure}[hbp]
\includegraphics[scale=1.0,angle=90]{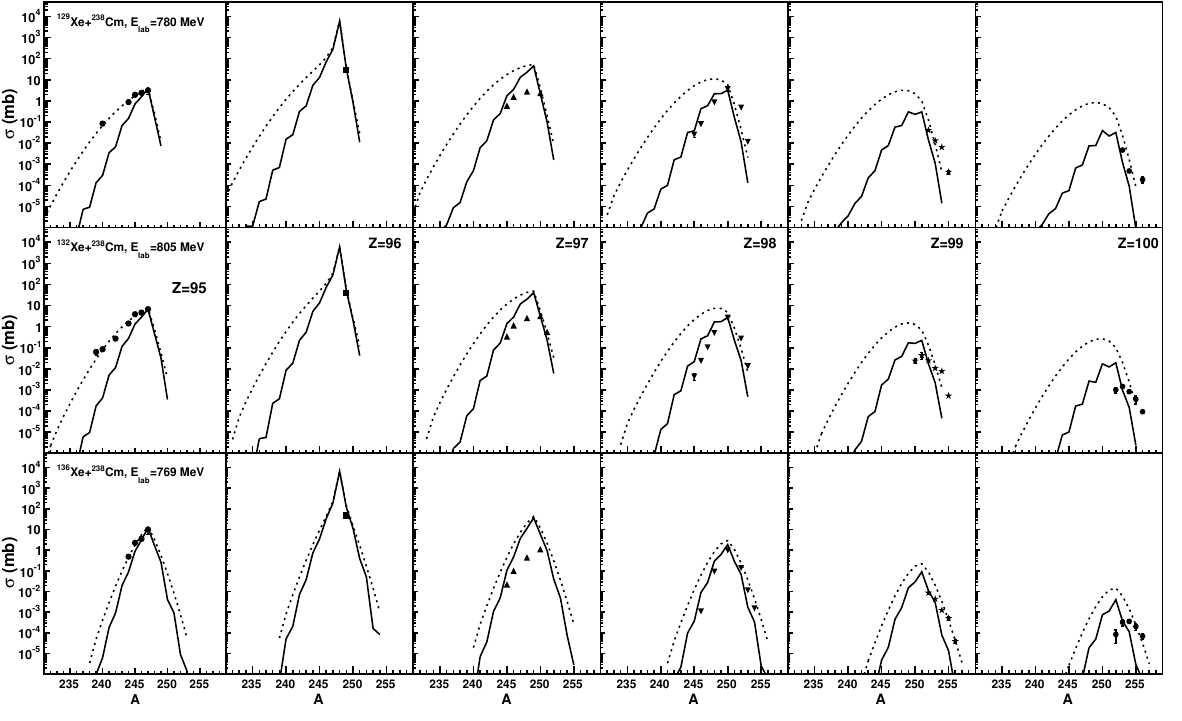}
\caption{Predicted (GRAZING-F) cross sections of surviving nuclei in the reaction $^{129}$Xe+$^{248}$Cm at $E_{lab}=780$ MeV, $^{132}$Xe+$^{248}$Cm at $E_{lab}=805$ MeV and $^{136}$Xe+$^{248}$Cm at $E_{lab}=769$ MeV (solid lines) compared to experimental data \cite{welch87,moody86} (solid symbols). Predicted primary product yields are shown as dotted lines.}
\label{fig_xe-cm248_data}
\end{figure}

Fig.~\ref{fig_xe-cm248_data} shows the comparison between experiment (solid symbols), the simulation of secondary (solid lines) and primary product yields (dotted lines). The agreement between prediction and measurement is quite reasonable. GRAZING-F seems to do a fair job in predicting the cross sections of $+2p$ transfer reactions but fails to reproduce the data for larger $p$ transfers and some below-target yields.

\subsection{The $^{136}$Xe+$^{249}$Cf reaction}

\begin{figure}[hbp]
\includegraphics[scale=0.9]{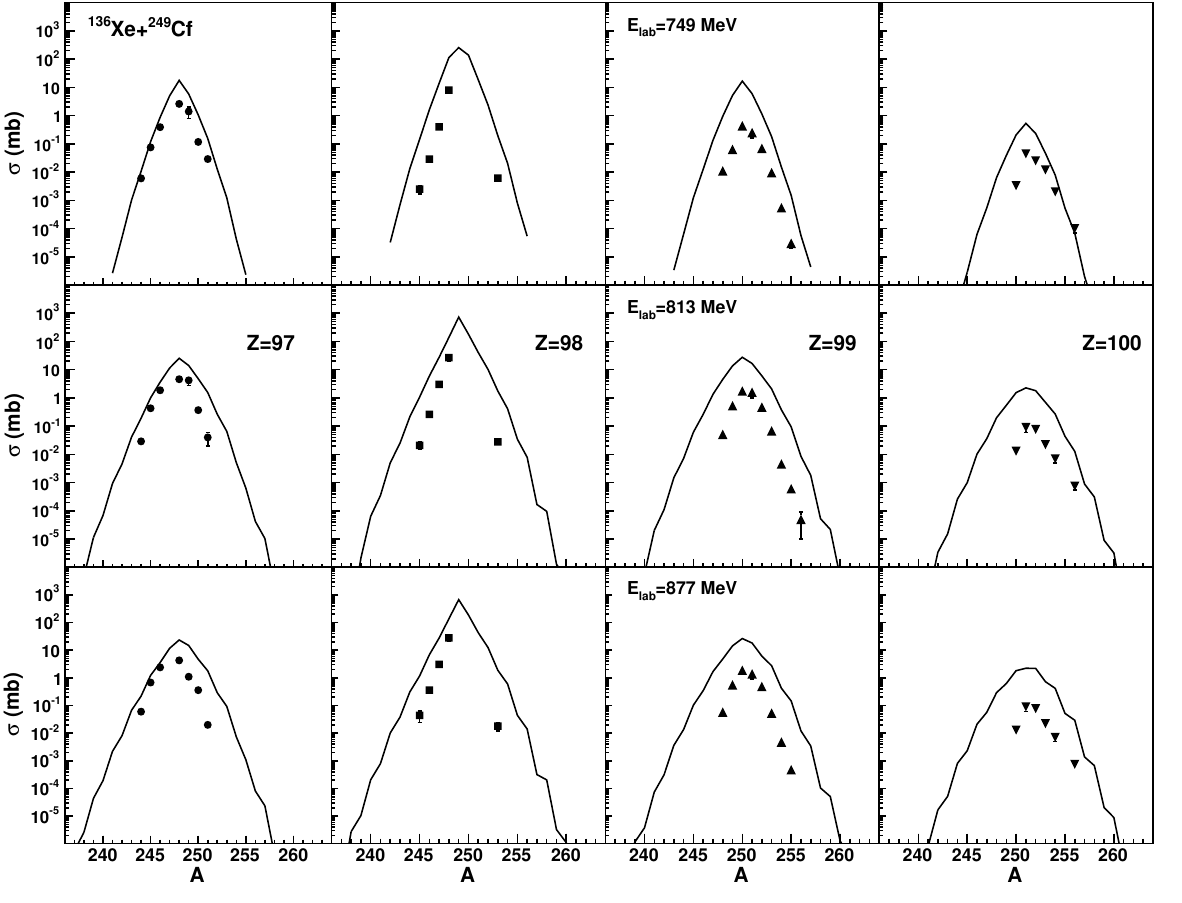}
\caption{Predicted (GRAZING-F) cross sections of surviving nuclei in the reaction $^{136}$Xe+$^{249}$Cf at $E_{lab}=749, 813, 877$ MeV (solid lines) compared to experimental data \cite{gregorich87} (solid symbols).}
\label{fig_xe136-cf249}
\end{figure}

The $^{136}$Xe+$^{249}$Cf reaction was measured with the intent to study the feasibility of using low-energy multi-nucleon transfer reactions to produce new actinide and transactinide isotopes \cite{gregorich87}. Fig.~\ref{fig_xe136-cf249} shows a comparison between experimental data and simulation with GRAZING-F for the three mid-target energies studied, $E_{lab}=749, 813, 877$ MeV, respectively. In this case, the simulations seem to predict the location of the maximum of the mass yields, but fail to predict the absolute values, overestimating the cross sections by an order of magnitude.

\subsection{The $^{136}$Xe+$^{244}$Pu reaction}

\begin{figure}[hbp]
\includegraphics[scale=0.7]{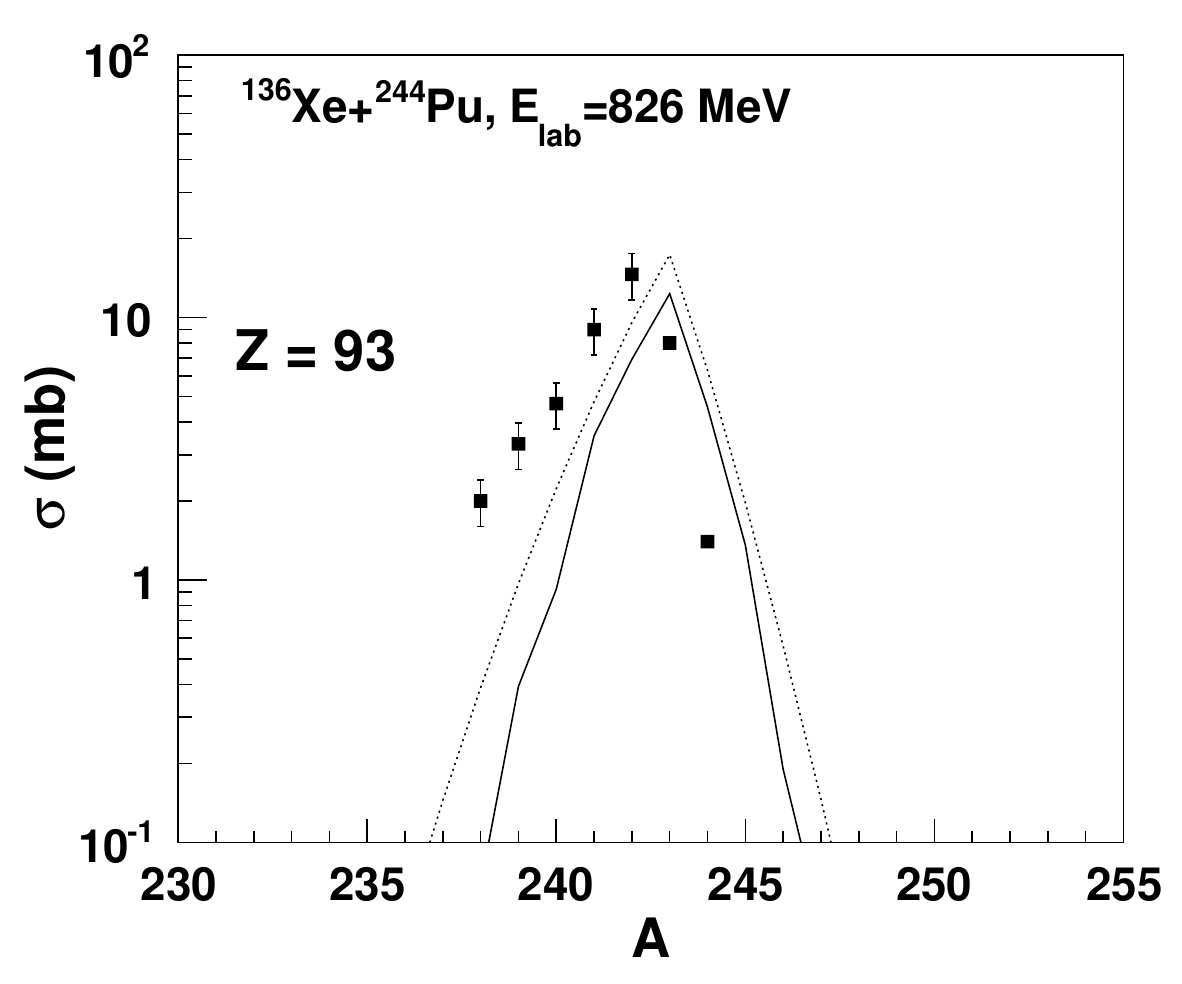}
\caption{Predicted (GRAZING-F) cross sections of surviving nuclei in the reaction $^{136}$Xe+$^{244}$Pu at $E_{lab}=826$ MeV (solid lines) compared to experimental data \cite{moody87} (solid symbols). Predicted primary product yields are shown as dotted lines.}
\label{fig_xe136-pu244}
\end{figure}

The reaction $^{136}$Xe+$^{244}$Pu at $E_{lab}=835$ MeV was used to produce and study the decay properties of the neutron-rich isotopes $^{243}$Np and $^{244}$Np \cite{moody87}. In Fig.~\ref{fig_xe136-pu244} we show the measured production cross section of Np isotopes compared to the predictions of GRAZING-F simulations. The simulations were done at energy $E_{lab}=826$ MeV, which we have estimated to be the mid-target energy. The predicted yield pattern is more neutron rich than the observed yield pattern but is similar in shape.

\subsection{The $^{86}$Kr+$^{248}$Cm reaction}

\begin{figure}[hbp]
\includegraphics[scale=1.0,angle=90]{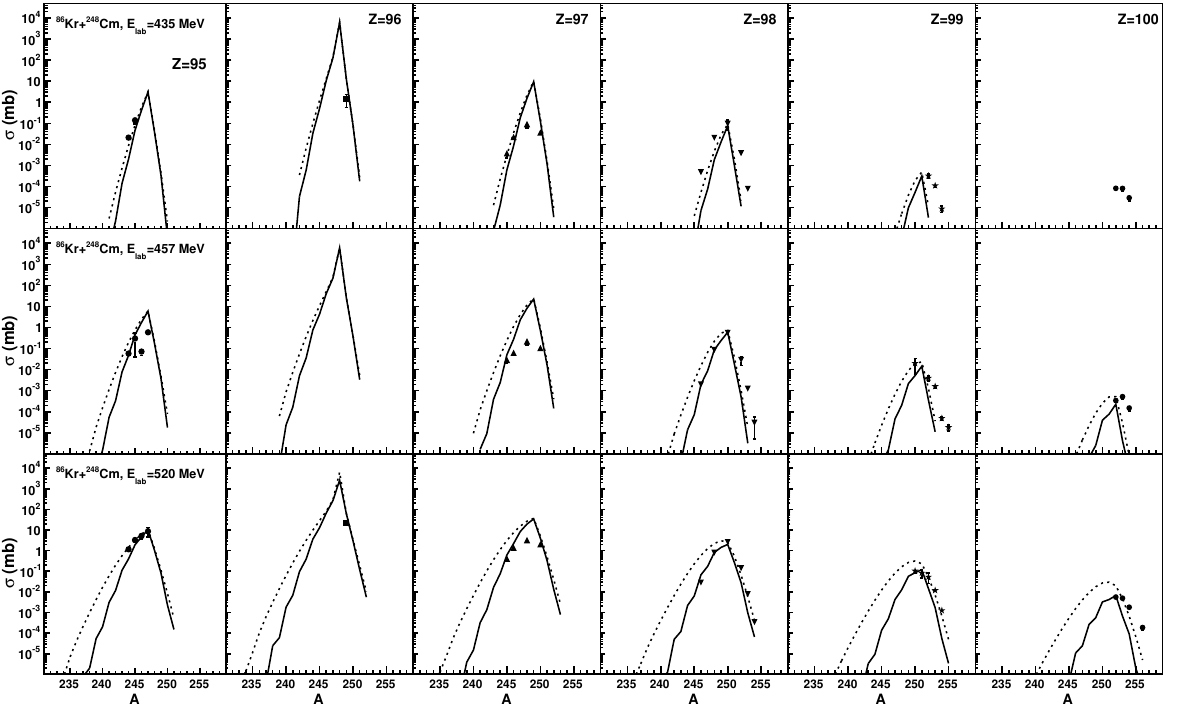}
\caption{Predicted (GRAZING-F) cross sections of surviving nuclei in the reaction $^{86}$Kr+$^{248}$Cm at $E_{lab}=435, 457$ and $520$ MeV (solid lines) compared to experimental data \cite{moody86} (solid symbols). Predicted primary product yields are shown as dotted lines.}
\label{fig_kr86-cm248}
\end{figure}

The $^{86}$Kr+$^{248}$Cm reaction was studied experimentally in the 1980s \cite{moody86}. Our simulations were done at $E_{lab}=435$ and $457$ MeV, corresponding to the entrance projectile energy, and $E_{lab}=520$ MeV, which we have estimated to be the mid-target energy for the reaction with entering energy of $546$ MeV. The former two energies are either below or at the interaction barriers (see Table~\ref{tab1}.) Fig.~\ref{fig_kr86-cm248} shows the comparison between experiment and simulations of secondary (solid lines) and primary product yields (dotted lines.) The observed yields are generally well represented by the GRAZING-F calculations.

\section{Predictions}

\subsection{The $^{136}$Xe+$^{208}$Pb reaction}

\begin{figure}[hbp]
\includegraphics[scale=0.8]{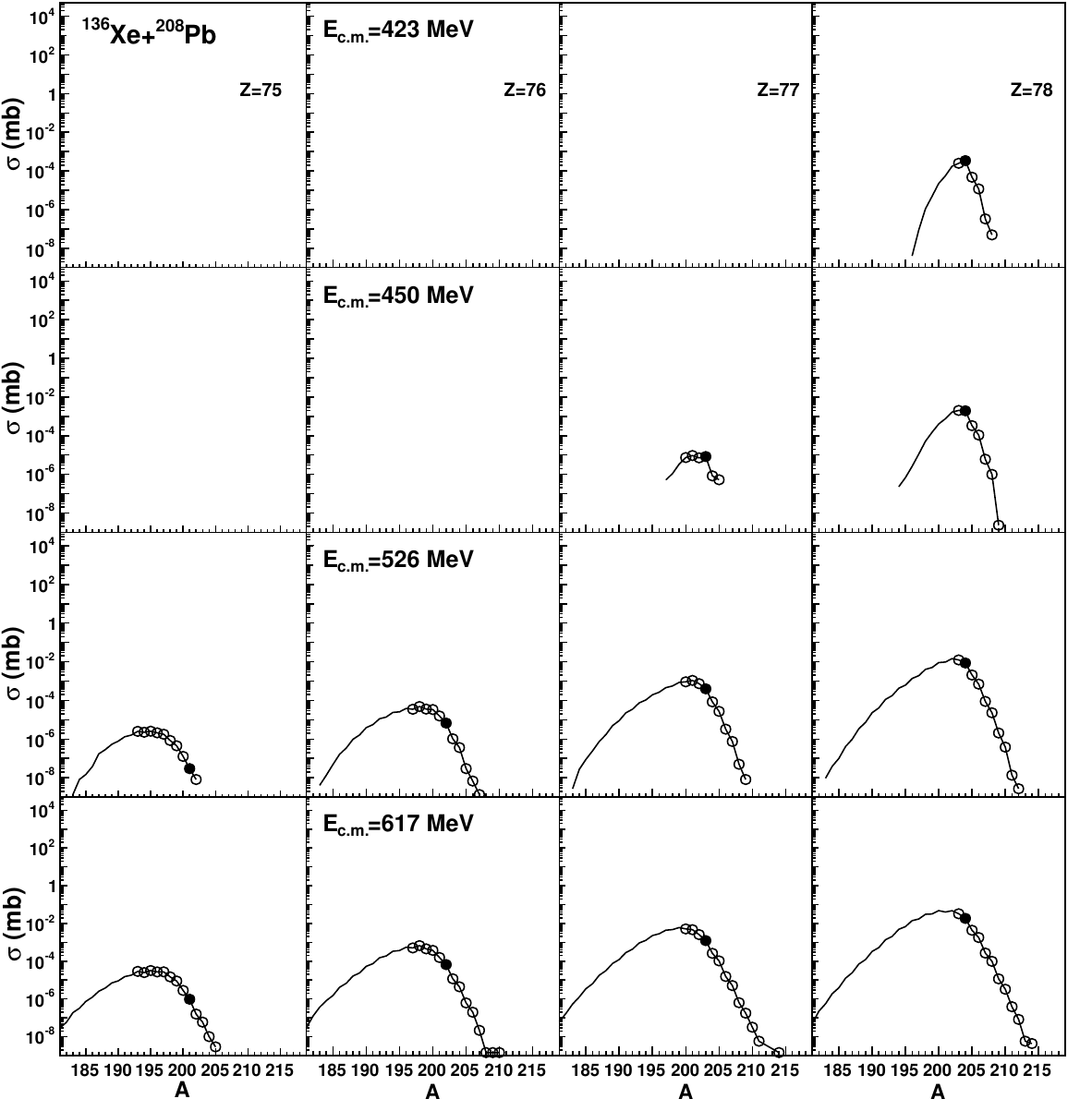}
\caption{Cross sections of surviving nuclei in the reaction $^{136}$Xe+$^{208}$Pb at $E_{c.m.}=423,450,526$ and $617$ MeV. The simulations (GRAZING-F) are shown as solid lines. Unknown isotopes are shown as open circles and unknown isotopes with $N=126$ are shown as solid circles.}
\label{fig_xe136-pb208}
\end{figure}

The study of the $^{136}$Xe+$^{208}$Pb reaction was first proposed by Zagrebaev and Greiner \cite{zaggy08prl} as a way of demonstrating how nuclear structure effects could be influencing the flow of nucleons in low-energy multi-nucleon transfer reactions towards both the $Z=82$ and $N=126$ closed shells. A dynamical model based on the multi-dimensional Langevin equations was used and the reaction was studied at $E_{c.m.}=450$ MeV. Mass-energy distributions of the $^{136}$Xe+$^{208}$Pb reaction have been measured recently with a double-arm time-of-flight spectrometer at $E_{c.m.}=423, 526$ and $617$ MeV \cite{kozulin12}. In Fig.~\ref{fig_xe136-pb208} we show the yields from the GRAZING-F simulations at energies $E_{c.m.}=423, 450, 526$ and $617$ MeV for transfers where unknown $N=126$ products are produced. (Unknown isotopes are plotted as open circles, while unknown $N=126$ isotopes are plotted as solid circles.) The transfer-fission cross section increases substantially in going from the lowest to the highest energy, from $\sim 10$ mb to $\sim 300$ mb, as can be intuitively expected due to the larger excitation energy of the primary products. The simulated yields for $^{203}_{77}$Ir and $^{204}_{78}$Pt peak at $E_{c.m.}\sim 750$ MeV, with cross sections of $1.7$ and $23$ $\mu$b, respectively. Assuming realistically a beam intensity of 100 pnA and a target thickness of 1 mg/cm$^2$, the production rates at this energy would be $3$ and $40$ s$^{-1}$, respectively. The range up to the interaction barrier is $2.75$ mg/cm$^2$. Hence, the maximum production rate at this energy and beam intensity would be $8$ and $110$ s$^{-1}$, respectively, assuming the cross section between $V_{int}$ and $E_{c.m.}$ varies slowly with energy. The simulations suggest that the $^{136}$Xe + $^{208}$Pb reaction can be an important source of new neutron-rich nuclei near the $N=126$ shell.

\subsection{The $^{136}$Xe+$^{198}$Pt reaction}

The $^{136}$Xe+$^{198}$Pt reaction at $E_{lab}=9$ MeV/A has been proposed as a $N=126$ ``factory'' based upon calculations using the GRAZING code without de-excitation by fission \cite{savard14}. If these predictions are correct, the properties of many unknown neutron-rich $N=126$ nuclei could be studied with intense $^{136}$Xe beams and thick $^{198}$Pt targets. Although this reaction has not been studied experimentally, we have performed simulations with GRAZING-F in case fission plays a role in the de-excitation of primary reaction products. We find that GRAZING-F (assuming $J=0 \hbar$) predicts a transfer-fission cross section of $\sim 30$ mb. Compared to the transfer cross section of $\sim 5$ b, fission does not seem to play a role if $J$ is low. Even if the transferred angular momentum is large, say $J=30 \hbar$, the largest angular momentum transferred predicted by GRAZING, the isotopic yields are essentially the same. Hence, fission competition nor angular momentum seem to play a role in this reaction.

In Table~\ref{tab3} we show the maximum production rates for $N=126$ isotopes by assuming a beam intensity of 1 p$\mu$A and a target thickness equivalent to the range from the entrance energy to the interaction barrier. The simulations suggest that the use of $^{198}$Pt as a $N=126$ ``factory'' is justified \cite{savard14} and may have a significant advantage over $^{208}$Pb, as the simulated transfer cross section in the former case may be a factor of two higher.

\subsection{The $^{144}$Xe+$^{248}$Cm reaction}

\begin{figure}[hbp]
\includegraphics[scale=1.0,angle=90]{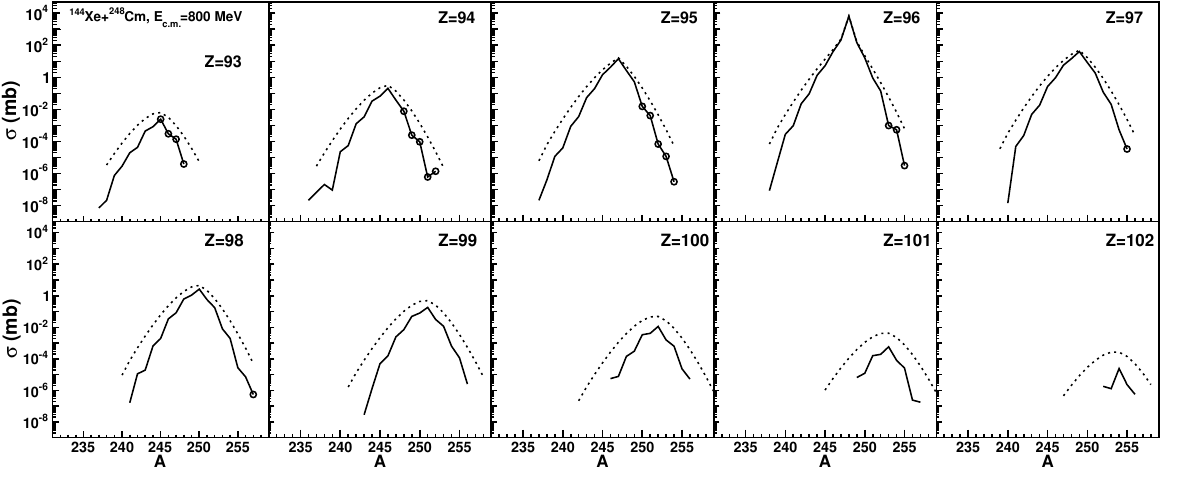}
\caption{Cross sections of surviving nuclei in the reaction $^{144}$Xe+$^{248}$Cm at $E_{lab}=800$ MeV. The predictions (GRAZING-F) are shown as solid lines. Unknown isotopes are shown as open circles. Predicted primary product yields are shown as dotted lines.}
\label{fig_pollarolo}
\end{figure}

The $^{144}$Xe+$^{248}$Cm reaction at $E_{lab}=800$ MeV has been proposed as an example reaction to be studied at EURISOL in a series of meetings and workshops \cite{pollarolo02,schadel10}. The presentations suggest that GRAZING predicts this reaction has production cross sections of the order of 0.1 mb for U, 1 mb for Np and Pu, and 10 mb for Am neutron-rich isotopes. However, these calculations did not consider neutron decay and thus represent yields of primary fragments only \cite{pollarolo}. Fig.~\ref{fig_pollarolo} shows the predictions of GRAZING-F. Unknown isotopes are shown as open circles. The present simulations predict cross sections of the order of $\mu$b to nb for the most neutron-rich unknown actinides. The transfer cross section is estimated to be $\sim 7$ b, whereas the transfer-fission cross section is estimated to be $\sim 160$ mb indicating that fission is not an important decay mode for the most n-rich products. GRAZING does not predict the production of $Z=92$ isotopes.

\begin{figure}[hbp]
\includegraphics[scale=1.1,angle=90]{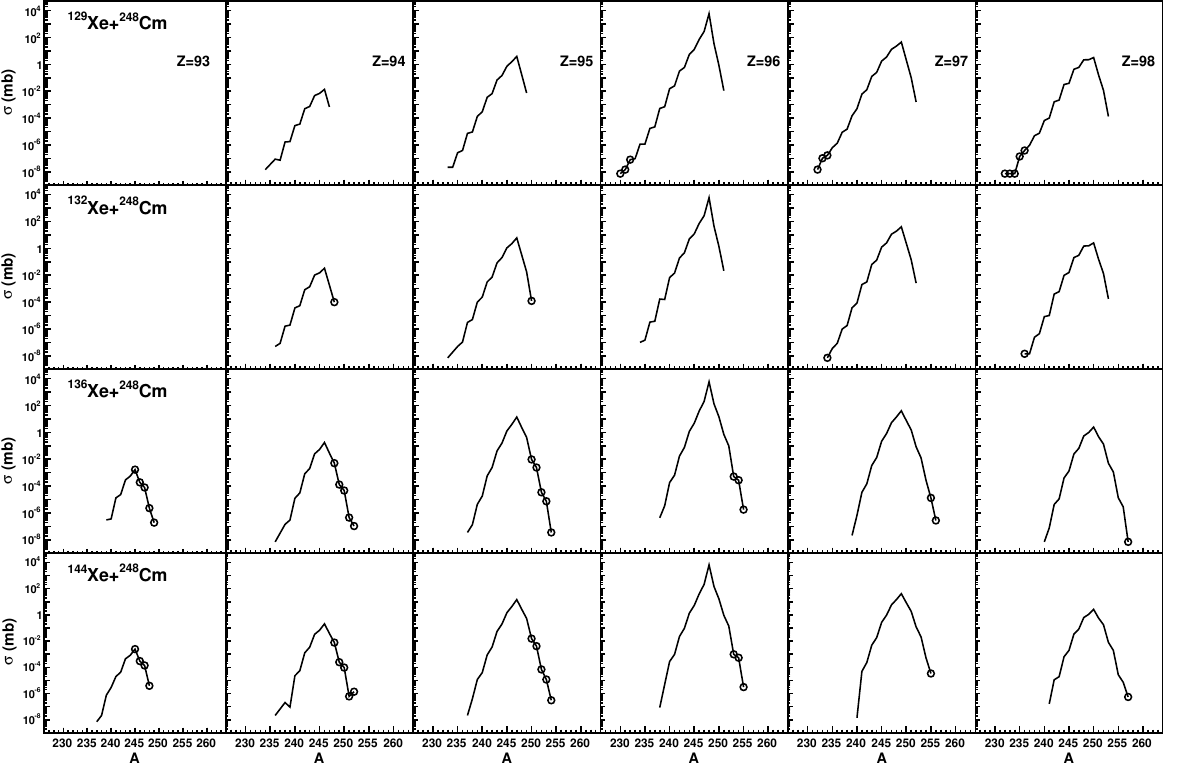}
\caption{Cross sections of surviving nuclei in the reaction $^{129,132,136,144}$Xe+$^{248}$Cm at $E_{c.m.}/V_int=1.05$. The predictions (GRAZING-F) are shown as solid lines. Unknown isotopes are shown as open circles.}
\label{fig_xe-cm248}
\end{figure}

The trend as a function of projectile mass is shown in Fig.~\ref{fig_xe-cm248}, where we plot the cross section of surviving nuclei of $-2p$, $-1p$, $0p$, $+1p$ and $+2p$ transfers for the  $^{129,132,136,144}$Xe+$^{248}$Cm at $E_{c.m.}/V_C=1.05$. Unknown isotopes are shown as open symbols. If we focus on cases that are well described by GRAZING-F ($+2p$ transfers), these simulations seem to indicate that a more neutron-rich projectile does indeed produce more neutron-rich products, but the advantage of going from the most neutron-rich stable Xe to a neutron-rich radioactive Xe projectile may not be as pronounced as claimed or hoped.  One notes furthermore that the projected intensities of $^{144}$Xe beams at modern radioactive beam facilities are a tiny fraction of the intensities of the stable Xe beams.

\subsection{The $^{94}$Kr+$^{248}$Cm reaction}

\begin{figure}[hbp]
\includegraphics[scale=1.0,angle=90]{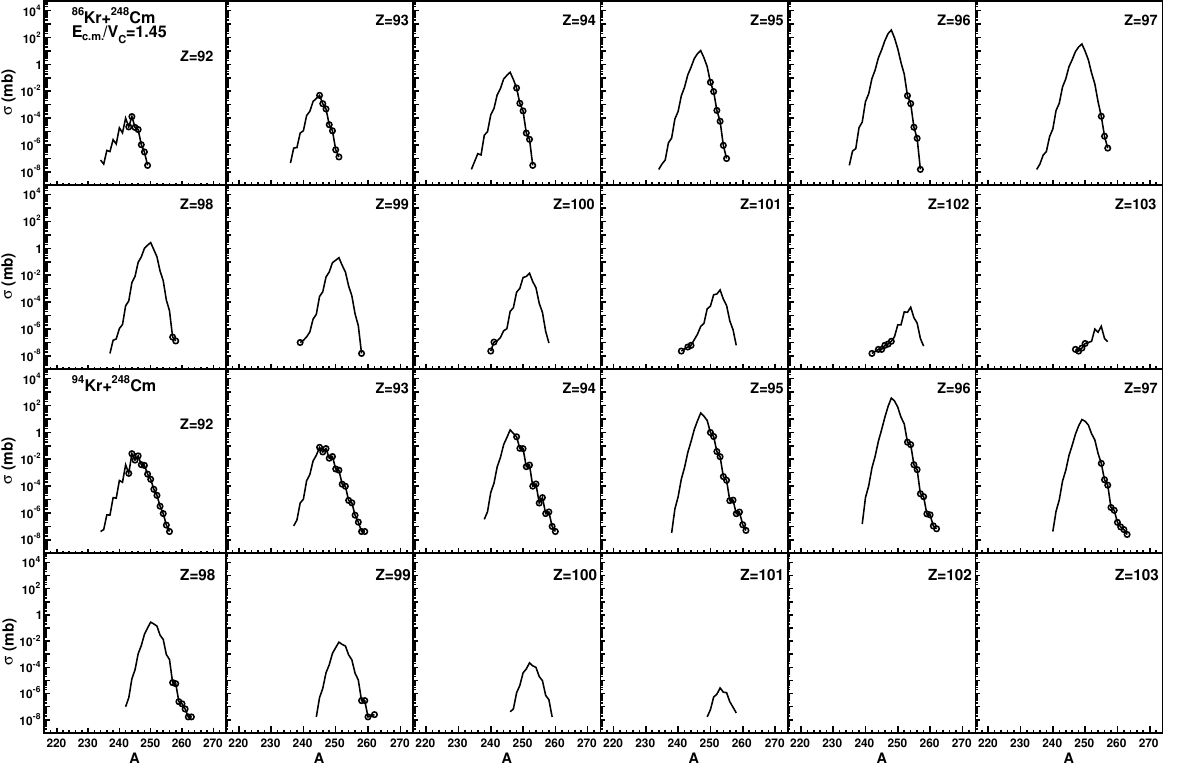}
\caption{Predicted (GRAZING-F) cross sections of surviving nuclei in the reaction $^{86,94}$Kr+$^{248}$Cm at $E_{c.m.}/V_{C}=1.45$. The predictions (GRAZING-F) are shown as solid lines. Unknown isotopes are shown as open circles.}
\label{fig_kr-cm248}
\end{figure}

The $^{94}$Kr+$^{248}$Cm reaction has been simulated at $E_{c.m.}/V_{C}=1.45$ and compared to the $^{86}$Kr+$^{248}$Cm reaction in Fig.~\ref{fig_kr-cm248}. Unknown isotopes are shown as open circles. The $^{94}$Kr simulations predict substantial cross sections for unknown neutron-rich actinide isotopes compared to $^{86}$Kr. For example, the predicted production cross section for $^{248}$Pu is $\sim 0.5$ mb in the $^{94}$Kr induced reaction, whereas the cross section is $\sim 0.02$ mb in the $^{86}$Kr induced reaction. Nonetheless, current $^{94}$Kr intensities are far too low for this reaction to be feasible to produce actinides. For example, the maximum production rate of $^{248}$Pu would be $\sim 15$ per year assuming the current CARIBU beam intensity estimate of $15$ s$^{-1}$.

The simulations also predict the absence of larger p transfers ($>+5p$) in the $^{94}$Kr induced reaction.

\subsection{The $^{238}$U+$^{249}$Bk reaction}

The $^{238}$U+$^{249}$Bk reaction has been suggested to be studied with a mass separator like the Fragment Mass Analyzer (FMA) at Argonne National Laboratory. The reason is that there is some evidence that large yields of transfer products could be observed close to $0^{\circ}$ \cite{loveland14}. If this is the case, the yields of short-lived neutron-rich actinides could be measured and the theory of Zagrebaev and Greiner \cite{zaggy06} could be tested. The particular choice, $^{238}$U+$^{249}$Bk, has been suggested because of convenience; $^{238}$U is the heaviest projectile accelerated by \uppercase{ATLAS} and a thin $^{249}$Bk target has recently become available.

\begin{figure}[hbp]
\includegraphics[scale=1.1,angle=90]{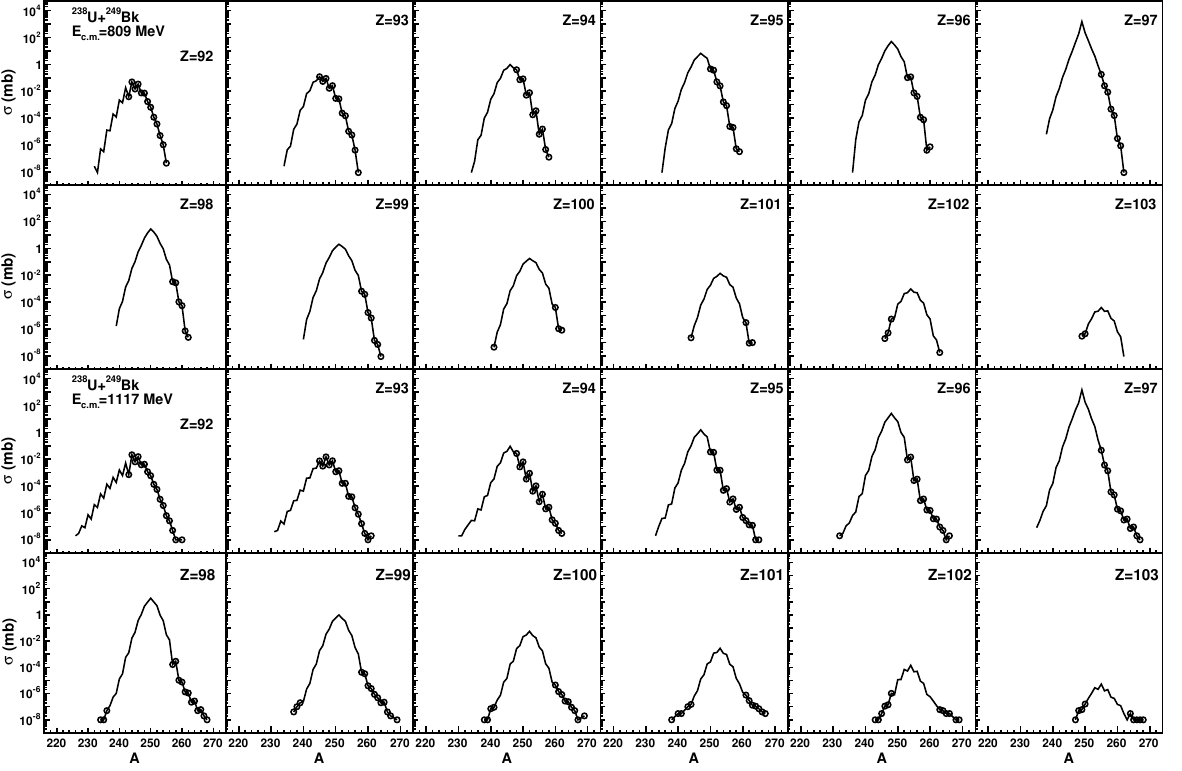}
\caption{Cross sections of surviving nuclei in the reaction $^{238}$U+$^{249}$Bk at $E_{c.m.}/V_{C}=1.05$ and $1.45$. The predictions (GRAZING-F) are shown as solid lines. Unknown isotopes are shown as open circles.}
\label{fig_u238-bk249}
\end{figure}

Fig.~\ref{fig_u238-bk249} shows the cross section of surviving nuclei predicted by GRAZING-F when $E_{c.m.}/V_C=1.05$ and $1.45$, respectively. Unknown isotopes are shown as open circles. The simulations predict substantial cross sections for unknown U and Pu isotopes at both energies, with the larger cross sections associated with the low-energy reaction due to fission competition. Table~\ref{tab5} shows the yields of unknown U and Pu isotopes assuming a beam intensity of 100 pnA, a target thickness of 0.3 mg/cm$^2$ and one day of irradiation. In the high-energy reaction GRAZING-F predicts the production of unknown neutron-rich Md, No and Lr isotopes with cross sections below 1 nb. Under the above assumptions, $^{261}$Md would have a yield of $\sim 30$, $^{261}$No $\sim 10$ and $^{263}$Lr $\simeq 1$ nuclei, respectively.

\section{Conclusions and discussion}

From the comparison of available experimental data with simulations of GRAZING-F we conclude that it is able to reproduce the yields of above-target products of $+1p$, $+2p$ and $+3p$ transfers reasonably well. The yields of $+1p$, $+2p$ and $+3p$ transfers in the  $^{238}$U+$^{248}$Cm reaction, for example, are exceptionally well reproduced. The yields of products involving larger proton transfers ($> +3p$) start to deviate substantially from experimental data primarily because GRAZING seems to underestimate the flow of neutrons. The usefulness of very neutron-rich radioactive beams, however intense, is predicted to be doubtful compared to neutron-rich stable beams, e.g. $^{86}$Kr and $^{136}$Xe, which are far more intense than any predicted intensities of the radioactive beams at planned radioactive beam facilities. GRAZING-F predicts substantial yields of unknown actinide isotopes under special conditions. For example, in the low energy U+Bk reaction, several unknown U and Pu isotopes could be produced with measurable yields. The production of unknown $N=126$ isotopes is predicted to be better accomplished by the use of the Xe+Pt reaction. The low transfer cross section and the high transfer-fission cross section associated with the Xe+Pb reaction makes it a less attractive candidate. Of course, these are predictions that must be verified by experiment. 

\begin{acknowledgments}
This work was supported in part by the Director, Office of Energy Research, Division of Nuclear Physics of the Office of High Energy and Nuclear Physics of the US Department of Energy under Grant DE-FG06-97ER41026.
\end{acknowledgments}

\bibliography{grazing}

\begin{thebibliography}{46}
\expandafter\ifx\csname natexlab\endcsname\relax\def\natexlab#1{#1}\fi
\expandafter\ifx\csname bibnamefont\endcsname\relax
  \def\bibnamefont#1{#1}\fi
\expandafter\ifx\csname bibfnamefont\endcsname\relax
  \def\bibfnamefont#1{#1}\fi
\expandafter\ifx\csname citenamefont\endcsname\relax
  \def\citenamefont#1{#1}\fi
\expandafter\ifx\csname url\endcsname\relax
  \def\url#1{\texttt{#1}}\fi
\expandafter\ifx\csname urlprefix\endcsname\relax\def\urlprefix{URL }\fi
\providecommand{\bibinfo}[2]{#2}
\providecommand{\eprint}[2][]{\url{#2}}

\bibitem[{\citenamefont{Schr\"oder and Huizenga}(1984)}]{schroder84}
\bibinfo{author}{\bibfnamefont{W.~U.} \bibnamefont{Schr\"oder}}
  \bibnamefont{and} \bibinfo{author}{\bibfnamefont{J.~R.}
  \bibnamefont{Huizenga}}, \emph{\bibinfo{title}{Treatise on Heavy-Ion
  Science}}, vol.~\bibinfo{volume}{2} (\bibinfo{publisher}{Plenum Press},
  \bibinfo{address}{New York}, \bibinfo{year}{1984}).

\bibitem[{\citenamefont{Volkov}(1989)}]{volkov89}
\bibinfo{author}{\bibfnamefont{V.~V.} \bibnamefont{Volkov}},
  \emph{\bibinfo{title}{Treatise on Heavy-Ion Science}},
  vol.~\bibinfo{volume}{8} (\bibinfo{publisher}{Plenum Press},
  \bibinfo{address}{New York}, \bibinfo{year}{1989}).

\bibitem[{\citenamefont{Sch\"adel et~al.}(1978)\citenamefont{Sch\"adel, Kratz,
  Ahrens, Br\"uchle, Franz, G\"aggeler, Warnecke, Wirth, Herrmann, Trautmann
  et~al.}}]{schadel78}
\bibinfo{author}{\bibfnamefont{M.}~\bibnamefont{Sch\"adel}},
  \bibinfo{author}{\bibfnamefont{J.~V.} \bibnamefont{Kratz}},
  \bibinfo{author}{\bibfnamefont{H.}~\bibnamefont{Ahrens}},
  \bibinfo{author}{\bibfnamefont{W.}~\bibnamefont{Br\"uchle}},
  \bibinfo{author}{\bibfnamefont{G.}~\bibnamefont{Franz}},
  \bibinfo{author}{\bibfnamefont{H.}~\bibnamefont{G\"aggeler}},
  \bibinfo{author}{\bibfnamefont{I.}~\bibnamefont{Warnecke}},
  \bibinfo{author}{\bibfnamefont{G.}~\bibnamefont{Wirth}},
  \bibinfo{author}{\bibfnamefont{G.}~\bibnamefont{Herrmann}},
  \bibinfo{author}{\bibfnamefont{N.}~\bibnamefont{Trautmann}},
  \bibnamefont{et~al.}, \bibinfo{journal}{Phys. Rev. Lett.}
  \textbf{\bibinfo{volume}{41}}, \bibinfo{pages}{469} (\bibinfo{year}{1978}),
  \urlprefix\url{http://link.aps.org/doi/10.1103/PhysRevLett.41.469}.

\bibitem[{\citenamefont{Sch\"adel et~al.}(1982)\citenamefont{Sch\"adel,
  Br\"uchle, G\"aggeler, Kratz, S\"ummerer, Wirth, Herrmann, Stakemann, Tittel,
  Trautmann et~al.}}]{schadel82}
\bibinfo{author}{\bibfnamefont{M.}~\bibnamefont{Sch\"adel}},
  \bibinfo{author}{\bibfnamefont{W.}~\bibnamefont{Br\"uchle}},
  \bibinfo{author}{\bibfnamefont{H.}~\bibnamefont{G\"aggeler}},
  \bibinfo{author}{\bibfnamefont{J.~V.} \bibnamefont{Kratz}},
  \bibinfo{author}{\bibfnamefont{K.}~\bibnamefont{S\"ummerer}},
  \bibinfo{author}{\bibfnamefont{G.}~\bibnamefont{Wirth}},
  \bibinfo{author}{\bibfnamefont{G.}~\bibnamefont{Herrmann}},
  \bibinfo{author}{\bibfnamefont{R.}~\bibnamefont{Stakemann}},
  \bibinfo{author}{\bibfnamefont{G.}~\bibnamefont{Tittel}},
  \bibinfo{author}{\bibfnamefont{N.}~\bibnamefont{Trautmann}},
  \bibnamefont{et~al.}, \bibinfo{journal}{Phys. Rev. Lett.}
  \textbf{\bibinfo{volume}{48}}, \bibinfo{pages}{852} (\bibinfo{year}{1982}),
  \urlprefix\url{http://link.aps.org/doi/10.1103/PhysRevLett.48.852}.

\bibitem[{\citenamefont{Lee et~al.}(1982)\citenamefont{Lee, von Gunten, Jacak,
  Nurmia, Liu, Luo, Seaborg, and Hoffman}}]{lee82}
\bibinfo{author}{\bibfnamefont{D.}~\bibnamefont{Lee}},
  \bibinfo{author}{\bibfnamefont{H.}~\bibnamefont{von Gunten}},
  \bibinfo{author}{\bibfnamefont{B.}~\bibnamefont{Jacak}},
  \bibinfo{author}{\bibfnamefont{M.}~\bibnamefont{Nurmia}},
  \bibinfo{author}{\bibfnamefont{Y.-f.} \bibnamefont{Liu}},
  \bibinfo{author}{\bibfnamefont{C.}~\bibnamefont{Luo}},
  \bibinfo{author}{\bibfnamefont{G.~T.} \bibnamefont{Seaborg}},
  \bibnamefont{and} \bibinfo{author}{\bibfnamefont{D.~C.}
  \bibnamefont{Hoffman}}, \bibinfo{journal}{Phys. Rev. C}
  \textbf{\bibinfo{volume}{25}}, \bibinfo{pages}{286} (\bibinfo{year}{1982}),
  \urlprefix\url{http://link.aps.org/doi/10.1103/PhysRevC.25.286}.

\bibitem[{\citenamefont{Moody et~al.}(1986)\citenamefont{Moody, Lee, Welch,
  Gregorich, Seaborg, Lougheed, and Hulet}}]{moody86}
\bibinfo{author}{\bibfnamefont{K.~J.} \bibnamefont{Moody}},
  \bibinfo{author}{\bibfnamefont{D.}~\bibnamefont{Lee}},
  \bibinfo{author}{\bibfnamefont{R.~B.} \bibnamefont{Welch}},
  \bibinfo{author}{\bibfnamefont{K.~E.} \bibnamefont{Gregorich}},
  \bibinfo{author}{\bibfnamefont{G.~T.} \bibnamefont{Seaborg}},
  \bibinfo{author}{\bibfnamefont{R.~W.} \bibnamefont{Lougheed}},
  \bibnamefont{and} \bibinfo{author}{\bibfnamefont{E.~K.} \bibnamefont{Hulet}},
  \bibinfo{journal}{Phys. Rev. C} \textbf{\bibinfo{volume}{33}},
  \bibinfo{pages}{1315} (\bibinfo{year}{1986}),
  \urlprefix\url{http://link.aps.org/doi/10.1103/PhysRevC.33.1315}.

\bibitem[{\citenamefont{Moody et~al.}(1987)\citenamefont{Moody, Br\"uchle,
  Br\"ugger, G\"aggeler, Haefner, Sch\"adel, S\"ummerer, Tetzlaff, Herrmann,
  Kaffrell et~al.}}]{moody87}
\bibinfo{author}{\bibfnamefont{K.}~\bibnamefont{Moody}},
  \bibinfo{author}{\bibfnamefont{W.}~\bibnamefont{Br\"uchle}},
  \bibinfo{author}{\bibfnamefont{M.}~\bibnamefont{Br\"ugger}},
  \bibinfo{author}{\bibfnamefont{H.}~\bibnamefont{G\"aggeler}},
  \bibinfo{author}{\bibfnamefont{B.}~\bibnamefont{Haefner}},
  \bibinfo{author}{\bibfnamefont{M.}~\bibnamefont{Sch\"adel}},
  \bibinfo{author}{\bibfnamefont{K.}~\bibnamefont{S\"ummerer}},
  \bibinfo{author}{\bibfnamefont{H.}~\bibnamefont{Tetzlaff}},
  \bibinfo{author}{\bibfnamefont{G.}~\bibnamefont{Herrmann}},
  \bibinfo{author}{\bibfnamefont{N.}~\bibnamefont{Kaffrell}},
  \bibnamefont{et~al.}, \bibinfo{journal}{Zeitschrift f\"ur Physik A Atomic
  Nuclei} \textbf{\bibinfo{volume}{328}}, \bibinfo{pages}{417}
  (\bibinfo{year}{1987}), ISSN \bibinfo{issn}{0939-7922},
  \urlprefix\url{http://dx.doi.org/10.1007/BF01289627}.

\bibitem[{\citenamefont{Welch et~al.}(1987)\citenamefont{Welch, Moody,
  Gregorich, Lee, and Seaborg}}]{welch87}
\bibinfo{author}{\bibfnamefont{R.~B.} \bibnamefont{Welch}},
  \bibinfo{author}{\bibfnamefont{K.~J.} \bibnamefont{Moody}},
  \bibinfo{author}{\bibfnamefont{K.~E.} \bibnamefont{Gregorich}},
  \bibinfo{author}{\bibfnamefont{D.}~\bibnamefont{Lee}}, \bibnamefont{and}
  \bibinfo{author}{\bibfnamefont{G.~T.} \bibnamefont{Seaborg}},
  \bibinfo{journal}{Phys. Rev. C} \textbf{\bibinfo{volume}{35}},
  \bibinfo{pages}{204} (\bibinfo{year}{1987}),
  \urlprefix\url{http://link.aps.org/doi/10.1103/PhysRevC.35.204}.

\bibitem[{\citenamefont{Gregorich et~al.}(1987)\citenamefont{Gregorich, Moody,
  Lee, Kot, Welch, Wilmarth, and Seaborg}}]{gregorich87}
\bibinfo{author}{\bibfnamefont{K.~E.} \bibnamefont{Gregorich}},
  \bibinfo{author}{\bibfnamefont{K.~J.} \bibnamefont{Moody}},
  \bibinfo{author}{\bibfnamefont{D.}~\bibnamefont{Lee}},
  \bibinfo{author}{\bibfnamefont{W.~K.} \bibnamefont{Kot}},
  \bibinfo{author}{\bibfnamefont{R.~B.} \bibnamefont{Welch}},
  \bibinfo{author}{\bibfnamefont{P.~A.} \bibnamefont{Wilmarth}},
  \bibnamefont{and} \bibinfo{author}{\bibfnamefont{G.~T.}
  \bibnamefont{Seaborg}}, \bibinfo{journal}{Phys. Rev. C}
  \textbf{\bibinfo{volume}{35}}, \bibinfo{pages}{2117} (\bibinfo{year}{1987}),
  \urlprefix\url{http://link.aps.org/doi/10.1103/PhysRevC.35.2117}.

\bibitem[{\citenamefont{Zagrebaev et~al.}(2006)\citenamefont{Zagrebaev,
  Oganessian, Itkis, and Greiner}}]{zaggy06}
\bibinfo{author}{\bibfnamefont{V.~I.} \bibnamefont{Zagrebaev}},
  \bibinfo{author}{\bibfnamefont{Y.~T.} \bibnamefont{Oganessian}},
  \bibinfo{author}{\bibfnamefont{M.~G.} \bibnamefont{Itkis}}, \bibnamefont{and}
  \bibinfo{author}{\bibfnamefont{W.}~\bibnamefont{Greiner}},
  \bibinfo{journal}{Phys. Rev. C} \textbf{\bibinfo{volume}{73}},
  \bibinfo{pages}{031602} (\bibinfo{year}{2006}),
  \urlprefix\url{http://link.aps.org/doi/10.1103/PhysRevC.73.031602}.

\bibitem[{\citenamefont{Zagrebaev and Greiner}(2008{\natexlab{a}})}]{zaggy08}
\bibinfo{author}{\bibfnamefont{V.}~\bibnamefont{Zagrebaev}} \bibnamefont{and}
  \bibinfo{author}{\bibfnamefont{W.}~\bibnamefont{Greiner}},
  \bibinfo{journal}{Phys. Rev. C} \textbf{\bibinfo{volume}{78}},
  \bibinfo{pages}{034610} (\bibinfo{year}{2008}{\natexlab{a}}),
  \urlprefix\url{http://link.aps.org/doi/10.1103/PhysRevC.78.034610}.

\bibitem[{\citenamefont{Zagrebaev and Greiner}(2013)}]{zagrebaev13}
\bibinfo{author}{\bibfnamefont{V.~I.} \bibnamefont{Zagrebaev}}
  \bibnamefont{and} \bibinfo{author}{\bibfnamefont{W.}~\bibnamefont{Greiner}},
  \bibinfo{journal}{Phys. Rev. C} \textbf{\bibinfo{volume}{87}},
  \bibinfo{pages}{034608} (\bibinfo{year}{2013}),
  \urlprefix\url{http://link.aps.org/doi/10.1103/PhysRevC.87.034608}.

\bibitem[{\citenamefont{Duellmann}()}]{chris14}
\bibinfo{author}{\bibfnamefont{C.~H.} \bibnamefont{Duellmann}},
  \bibinfo{note}{private communication.}

\bibitem[{\citenamefont{Winther}(1994)}]{winther}
\bibinfo{author}{\bibfnamefont{A.}~\bibnamefont{Winther}},
  \bibinfo{journal}{Nuclear Physics A} \textbf{\bibinfo{volume}{572}},
  \bibinfo{pages}{191 } (\bibinfo{year}{1994}), ISSN \bibinfo{issn}{0375-9474},
  \urlprefix\url{http://www.sciencedirect.com/science/article/pii/0375947494904308}.

\bibitem[{\citenamefont{Winther}(1995)}]{winther2}
\bibinfo{author}{\bibfnamefont{A.}~\bibnamefont{Winther}},
  \bibinfo{journal}{Nuclear Physics A} \textbf{\bibinfo{volume}{594}},
  \bibinfo{pages}{203 } (\bibinfo{year}{1995}), ISSN \bibinfo{issn}{0375-9474},
  \urlprefix\url{http://www.sciencedirect.com/science/article/pii/037594749500374A}.

\bibitem[{\citenamefont{Winther}()}]{grazing}
\bibinfo{author}{\bibfnamefont{A.}~\bibnamefont{Winther}},
  \bibinfo{note}{\uppercase{GRAZING} version 9
  (http://personalpages.to.infn.it/$\sim$nanni/grazing/)}.

\bibitem[{\citenamefont{Dasso et~al.}(1994)\citenamefont{Dasso, Pollarolo, and
  Winther}}]{dasso94}
\bibinfo{author}{\bibfnamefont{C.~H.} \bibnamefont{Dasso}},
  \bibinfo{author}{\bibfnamefont{G.}~\bibnamefont{Pollarolo}},
  \bibnamefont{and} \bibinfo{author}{\bibfnamefont{A.}~\bibnamefont{Winther}},
  \bibinfo{journal}{Phys. Rev. Lett.} \textbf{\bibinfo{volume}{73}},
  \bibinfo{pages}{1907} (\bibinfo{year}{1994}),
  \urlprefix\url{http://link.aps.org/doi/10.1103/PhysRevLett.73.1907}.

\bibitem[{\citenamefont{Corradi et~al.}(1996)\citenamefont{Corradi, He,
  Ackermann, Stefanini, Pisent, Beghini, Montagnoli, Scarlassara, Segato,
  Pollarolo et~al.}}]{corradi96}
\bibinfo{author}{\bibfnamefont{L.}~\bibnamefont{Corradi}},
  \bibinfo{author}{\bibfnamefont{J.~H.} \bibnamefont{He}},
  \bibinfo{author}{\bibfnamefont{D.}~\bibnamefont{Ackermann}},
  \bibinfo{author}{\bibfnamefont{A.~M.} \bibnamefont{Stefanini}},
  \bibinfo{author}{\bibfnamefont{A.}~\bibnamefont{Pisent}},
  \bibinfo{author}{\bibfnamefont{S.}~\bibnamefont{Beghini}},
  \bibinfo{author}{\bibfnamefont{G.}~\bibnamefont{Montagnoli}},
  \bibinfo{author}{\bibfnamefont{F.}~\bibnamefont{Scarlassara}},
  \bibinfo{author}{\bibfnamefont{G.~F.} \bibnamefont{Segato}},
  \bibinfo{author}{\bibfnamefont{G.}~\bibnamefont{Pollarolo}},
  \bibnamefont{et~al.}, \bibinfo{journal}{Phys. Rev. C}
  \textbf{\bibinfo{volume}{54}}, \bibinfo{pages}{201} (\bibinfo{year}{1996}),
  \urlprefix\url{http://link.aps.org/doi/10.1103/PhysRevC.54.201}.

\bibitem[{\citenamefont{Corradi et~al.}(1997)\citenamefont{Corradi, Stefanini,
  He, Beghini, Montagnoli, Scarlassara, Segato, Pollarolo, and
  Dasso}}]{corradi97}
\bibinfo{author}{\bibfnamefont{L.}~\bibnamefont{Corradi}},
  \bibinfo{author}{\bibfnamefont{A.~M.} \bibnamefont{Stefanini}},
  \bibinfo{author}{\bibfnamefont{J.~H.} \bibnamefont{He}},
  \bibinfo{author}{\bibfnamefont{S.}~\bibnamefont{Beghini}},
  \bibinfo{author}{\bibfnamefont{G.}~\bibnamefont{Montagnoli}},
  \bibinfo{author}{\bibfnamefont{F.}~\bibnamefont{Scarlassara}},
  \bibinfo{author}{\bibfnamefont{G.~F.} \bibnamefont{Segato}},
  \bibinfo{author}{\bibfnamefont{G.}~\bibnamefont{Pollarolo}},
  \bibnamefont{and} \bibinfo{author}{\bibfnamefont{C.~H.} \bibnamefont{Dasso}},
  \bibinfo{journal}{Phys. Rev. C} \textbf{\bibinfo{volume}{56}},
  \bibinfo{pages}{938} (\bibinfo{year}{1997}),
  \urlprefix\url{http://link.aps.org/doi/10.1103/PhysRevC.56.938}.

\bibitem[{\citenamefont{Corradi et~al.}(1999)\citenamefont{Corradi, Stefanini,
  Lin, Beghini, Montagnoli, Scarlassara, Pollarolo, and Winther}}]{corradi99}
\bibinfo{author}{\bibfnamefont{L.}~\bibnamefont{Corradi}},
  \bibinfo{author}{\bibfnamefont{A.~M.} \bibnamefont{Stefanini}},
  \bibinfo{author}{\bibfnamefont{C.~J.} \bibnamefont{Lin}},
  \bibinfo{author}{\bibfnamefont{S.}~\bibnamefont{Beghini}},
  \bibinfo{author}{\bibfnamefont{G.}~\bibnamefont{Montagnoli}},
  \bibinfo{author}{\bibfnamefont{F.}~\bibnamefont{Scarlassara}},
  \bibinfo{author}{\bibfnamefont{G.}~\bibnamefont{Pollarolo}},
  \bibnamefont{and} \bibinfo{author}{\bibfnamefont{A.}~\bibnamefont{Winther}},
  \bibinfo{journal}{Phys. Rev. C} \textbf{\bibinfo{volume}{59}},
  \bibinfo{pages}{261} (\bibinfo{year}{1999}),
  \urlprefix\url{http://link.aps.org/doi/10.1103/PhysRevC.59.261}.

\bibitem[{\citenamefont{Corradi et~al.}(2001)\citenamefont{Corradi, Vinodkumar,
  Stefanini, Ackermann, Trotta, Beghini, Montagnoli, Scarlassara, Pollarolo,
  Cerutti et~al.}}]{corradi01}
\bibinfo{author}{\bibfnamefont{L.}~\bibnamefont{Corradi}},
  \bibinfo{author}{\bibfnamefont{A.~M.} \bibnamefont{Vinodkumar}},
  \bibinfo{author}{\bibfnamefont{A.~M.} \bibnamefont{Stefanini}},
  \bibinfo{author}{\bibfnamefont{D.}~\bibnamefont{Ackermann}},
  \bibinfo{author}{\bibfnamefont{M.}~\bibnamefont{Trotta}},
  \bibinfo{author}{\bibfnamefont{S.}~\bibnamefont{Beghini}},
  \bibinfo{author}{\bibfnamefont{G.}~\bibnamefont{Montagnoli}},
  \bibinfo{author}{\bibfnamefont{F.}~\bibnamefont{Scarlassara}},
  \bibinfo{author}{\bibfnamefont{G.}~\bibnamefont{Pollarolo}},
  \bibinfo{author}{\bibfnamefont{F.}~\bibnamefont{Cerutti}},
  \bibnamefont{et~al.}, \bibinfo{journal}{Phys. Rev. C}
  \textbf{\bibinfo{volume}{63}}, \bibinfo{pages}{021601}
  (\bibinfo{year}{2001}),
  \urlprefix\url{http://link.aps.org/doi/10.1103/PhysRevC.63.021601}.

\bibitem[{\citenamefont{Montagnoli et~al.}(2002)\citenamefont{Montagnoli,
  Beghini, Scarlassara, Stefanini, Corradi, Lin, Pollarolo, and
  Winther}}]{montagnoli02}
\bibinfo{author}{\bibfnamefont{G.}~\bibnamefont{Montagnoli}},
  \bibinfo{author}{\bibfnamefont{S.}~\bibnamefont{Beghini}},
  \bibinfo{author}{\bibfnamefont{F.}~\bibnamefont{Scarlassara}},
  \bibinfo{author}{\bibfnamefont{A.}~\bibnamefont{Stefanini}},
  \bibinfo{author}{\bibfnamefont{L.}~\bibnamefont{Corradi}},
  \bibinfo{author}{\bibfnamefont{C.}~\bibnamefont{Lin}},
  \bibinfo{author}{\bibfnamefont{G.}~\bibnamefont{Pollarolo}},
  \bibnamefont{and} \bibinfo{author}{\bibfnamefont{A.}~\bibnamefont{Winther}},
  \bibinfo{journal}{The European Physical Journal A - Hadrons and Nuclei}
  \textbf{\bibinfo{volume}{15}}, \bibinfo{pages}{351} (\bibinfo{year}{2002}),
  ISSN \bibinfo{issn}{1434-6001},
  \urlprefix\url{http://dx.doi.org/10.1140/epja/i2002-10034-8}.

\bibitem[{\citenamefont{Corradi
  et~al.}(2002{\natexlab{a}})\citenamefont{Corradi, Vinodkumar, Stefanini,
  Fioretto, Prete, Beghini, Montagnoli, Scarlassara, Pollarolo, Cerutti
  et~al.}}]{corradi02a}
\bibinfo{author}{\bibfnamefont{L.}~\bibnamefont{Corradi}},
  \bibinfo{author}{\bibfnamefont{A.~M.} \bibnamefont{Vinodkumar}},
  \bibinfo{author}{\bibfnamefont{A.~M.} \bibnamefont{Stefanini}},
  \bibinfo{author}{\bibfnamefont{E.}~\bibnamefont{Fioretto}},
  \bibinfo{author}{\bibfnamefont{G.}~\bibnamefont{Prete}},
  \bibinfo{author}{\bibfnamefont{S.}~\bibnamefont{Beghini}},
  \bibinfo{author}{\bibfnamefont{G.}~\bibnamefont{Montagnoli}},
  \bibinfo{author}{\bibfnamefont{F.}~\bibnamefont{Scarlassara}},
  \bibinfo{author}{\bibfnamefont{G.}~\bibnamefont{Pollarolo}},
  \bibinfo{author}{\bibfnamefont{F.}~\bibnamefont{Cerutti}},
  \bibnamefont{et~al.}, \bibinfo{journal}{Phys. Rev. C}
  \textbf{\bibinfo{volume}{66}}, \bibinfo{pages}{024606}
  (\bibinfo{year}{2002}{\natexlab{a}}),
  \urlprefix\url{http://link.aps.org/doi/10.1103/PhysRevC.66.024606}.

\bibitem[{\citenamefont{Corradi
  et~al.}(2002{\natexlab{b}})\citenamefont{Corradi, Stefanini, Vinodkumar,
  Beghini, Montagnoli, Scarlassara, and Pollarolo}}]{corradi02}
\bibinfo{author}{\bibfnamefont{L.}~\bibnamefont{Corradi}},
  \bibinfo{author}{\bibfnamefont{A.}~\bibnamefont{Stefanini}},
  \bibinfo{author}{\bibfnamefont{A.}~\bibnamefont{Vinodkumar}},
  \bibinfo{author}{\bibfnamefont{S.}~\bibnamefont{Beghini}},
  \bibinfo{author}{\bibfnamefont{G.}~\bibnamefont{Montagnoli}},
  \bibinfo{author}{\bibfnamefont{F.}~\bibnamefont{Scarlassara}},
  \bibnamefont{and}
  \bibinfo{author}{\bibfnamefont{G.}~\bibnamefont{Pollarolo}},
  \bibinfo{journal}{Nuclear Physics A} \textbf{\bibinfo{volume}{701}},
  \bibinfo{pages}{109 } (\bibinfo{year}{2002}{\natexlab{b}}), ISSN
  \bibinfo{issn}{0375-9474}, \bibinfo{note}{5th International Conference on
  Radioactive Nuclear Beams},
  \urlprefix\url{http://www.sciencedirect.com/science/article/pii/S0375947401015573}.

\bibitem[{\citenamefont{Corradi et~al.}(2013)\citenamefont{Corradi, Szilner,
  Pollarolo, Montanari, Fioretto, Stefanini, Valiente-Dobón, Farnea,
  Michelagnoli, Montagnoli et~al.}}]{corradi13}
\bibinfo{author}{\bibfnamefont{L.}~\bibnamefont{Corradi}},
  \bibinfo{author}{\bibfnamefont{S.}~\bibnamefont{Szilner}},
  \bibinfo{author}{\bibfnamefont{G.}~\bibnamefont{Pollarolo}},
  \bibinfo{author}{\bibfnamefont{D.}~\bibnamefont{Montanari}},
  \bibinfo{author}{\bibfnamefont{E.}~\bibnamefont{Fioretto}},
  \bibinfo{author}{\bibfnamefont{A.}~\bibnamefont{Stefanini}},
  \bibinfo{author}{\bibfnamefont{J.}~\bibnamefont{Valiente-Dobón}},
  \bibinfo{author}{\bibfnamefont{E.}~\bibnamefont{Farnea}},
  \bibinfo{author}{\bibfnamefont{C.}~\bibnamefont{Michelagnoli}},
  \bibinfo{author}{\bibfnamefont{G.}~\bibnamefont{Montagnoli}},
  \bibnamefont{et~al.}, \bibinfo{journal}{Nuclear Instruments and Methods in
  Physics Research Section B: Beam Interactions with Materials and Atoms}
  \textbf{\bibinfo{volume}{317, Part B}}, \bibinfo{pages}{743 }
  (\bibinfo{year}{2013}), ISSN \bibinfo{issn}{0168-583X},
  \bibinfo{note}{\{XVIth\} International Conference on ElectroMagnetic Isotope
  Separators and Techniques Related to their Applications, December 2–7, 2012
  at Matsue, Japan},
  \urlprefix\url{http://www.sciencedirect.com/science/article/pii/S0168583X13006976}.

\bibitem[{\citenamefont{Corradi et~al.}(2009)\citenamefont{Corradi, Pollarolo,
  and Szilner}}]{corradi09}
\bibinfo{author}{\bibfnamefont{L.}~\bibnamefont{Corradi}},
  \bibinfo{author}{\bibfnamefont{G.}~\bibnamefont{Pollarolo}},
  \bibnamefont{and} \bibinfo{author}{\bibfnamefont{S.}~\bibnamefont{Szilner}},
  \bibinfo{journal}{Journal of Physics G: Nuclear and Particle Physics}
  \textbf{\bibinfo{volume}{36}}, \bibinfo{pages}{113101}
  (\bibinfo{year}{2009}),
  \urlprefix\url{http://stacks.iop.org/0954-3899/36/i=11/a=113101}.

\bibitem[{\citenamefont{Schmidt et~al.}(1978)\citenamefont{Schmidt, Toneev, and
  Wolschin}}]{schmidt78}
\bibinfo{author}{\bibfnamefont{R.}~\bibnamefont{Schmidt}},
  \bibinfo{author}{\bibfnamefont{V.}~\bibnamefont{Toneev}}, \bibnamefont{and}
  \bibinfo{author}{\bibfnamefont{G.}~\bibnamefont{Wolschin}},
  \bibinfo{journal}{Nuclear Physics A} \textbf{\bibinfo{volume}{311}},
  \bibinfo{pages}{247 } (\bibinfo{year}{1978}), ISSN \bibinfo{issn}{0375-9474},
  \urlprefix\url{http://www.sciencedirect.com/science/article/pii/0375947478905134}.

\bibitem[{\citenamefont{Farra}(1996)}]{farra96}
\bibinfo{author}{\bibfnamefont{A.~A.} \bibnamefont{Farra}},
  \bibinfo{journal}{Canadian Journal of Physics} \textbf{\bibinfo{volume}{74}},
  \bibinfo{pages}{150} (\bibinfo{year}{1996}),
  \eprint{http://dx.doi.org/10.1139/p96-024},
  \urlprefix\url{http://dx.doi.org/10.1139/p96-024}.

\bibitem[{\citenamefont{Adamian et~al.}(2005)\citenamefont{Adamian, Antonenko,
  and Zubov}}]{adamian05}
\bibinfo{author}{\bibfnamefont{G.~G.} \bibnamefont{Adamian}},
  \bibinfo{author}{\bibfnamefont{N.~V.} \bibnamefont{Antonenko}},
  \bibnamefont{and} \bibinfo{author}{\bibfnamefont{A.~S.} \bibnamefont{Zubov}},
  \bibinfo{journal}{Phys. Rev. C} \textbf{\bibinfo{volume}{71}},
  \bibinfo{pages}{034603} (\bibinfo{year}{2005}),
  \urlprefix\url{http://link.aps.org/doi/10.1103/PhysRevC.71.034603}.

\bibitem[{\citenamefont{Kedziora and Simenel}(2010)}]{kedziora10}
\bibinfo{author}{\bibfnamefont{D.~J.} \bibnamefont{Kedziora}} \bibnamefont{and}
  \bibinfo{author}{\bibfnamefont{C.}~\bibnamefont{Simenel}},
  \bibinfo{journal}{Phys. Rev. C} \textbf{\bibinfo{volume}{81}},
  \bibinfo{pages}{044613} (\bibinfo{year}{2010}),
  \urlprefix\url{http://link.aps.org/doi/10.1103/PhysRevC.81.044613}.

\bibitem[{\citenamefont{Vandenbosch and Huizenga}(1973)}]{vandenbosch}
\bibinfo{author}{\bibfnamefont{R.}~\bibnamefont{Vandenbosch}} \bibnamefont{and}
  \bibinfo{author}{\bibfnamefont{J.}~\bibnamefont{Huizenga}},
  \emph{\bibinfo{title}{Nuclear Fission}} (\bibinfo{publisher}{Academic Press},
  \bibinfo{address}{New York}, \bibinfo{year}{1973}).

\bibitem[{\citenamefont{Myers and \ifmmode \acute{S}\else
  \'{S}\fi{}wia\ifmmode~\mbox{\c{}}\else \c{}\fi{}tecki}(1999)}]{myers99}
\bibinfo{author}{\bibfnamefont{W.~D.} \bibnamefont{Myers}} \bibnamefont{and}
  \bibinfo{author}{\bibfnamefont{W.~J.} \bibnamefont{\ifmmode \acute{S}\else
  \'{S}\fi{}wia\ifmmode~\mbox{\c{}}\else \c{}\fi{}tecki}},
  \bibinfo{journal}{Phys. Rev. C} \textbf{\bibinfo{volume}{60}},
  \bibinfo{pages}{014606} (\bibinfo{year}{1999}),
  \urlprefix\url{http://link.aps.org/doi/10.1103/PhysRevC.60.014606}.

\bibitem[{\citenamefont{Moller et~al.}(1995)\citenamefont{Moller, Nix, Myers,
  and Swiatecki}}]{moller95}
\bibinfo{author}{\bibfnamefont{P.}~\bibnamefont{Moller}},
  \bibinfo{author}{\bibfnamefont{J.}~\bibnamefont{Nix}},
  \bibinfo{author}{\bibfnamefont{W.}~\bibnamefont{Myers}}, \bibnamefont{and}
  \bibinfo{author}{\bibfnamefont{W.}~\bibnamefont{Swiatecki}},
  \bibinfo{journal}{Atomic Data and Nuclear Data Tables}
  \textbf{\bibinfo{volume}{59}}, \bibinfo{pages}{185 } (\bibinfo{year}{1995}),
  ISSN \bibinfo{issn}{0092-640X},
  \urlprefix\url{http://www.sciencedirect.com/science/article/pii/S0092640X85710029}.

\bibitem[{\citenamefont{Sierk}(1986)}]{sierk86}
\bibinfo{author}{\bibfnamefont{A.~J.} \bibnamefont{Sierk}},
  \bibinfo{journal}{Phys. Rev. C} \textbf{\bibinfo{volume}{33}},
  \bibinfo{pages}{2039} (\bibinfo{year}{1986}),
  \urlprefix\url{http://link.aps.org/doi/10.1103/PhysRevC.33.2039}.

\bibitem[{\citenamefont{Ignatyuk et~al.}(1976)\citenamefont{Ignatyuk, Itkis,
  Okolovich, Smirenkin, and Tishin}}]{ignatyuk75}
\bibinfo{author}{\bibfnamefont{A.}~\bibnamefont{Ignatyuk}},
  \bibinfo{author}{\bibfnamefont{M.}~\bibnamefont{Itkis}},
  \bibinfo{author}{\bibfnamefont{V.}~\bibnamefont{Okolovich}},
  \bibinfo{author}{\bibfnamefont{G.}~\bibnamefont{Smirenkin}},
  \bibnamefont{and} \bibinfo{author}{\bibfnamefont{A.}~\bibnamefont{Tishin}},
  \bibinfo{journal}{Sov. J. Nucl. Phys.} \textbf{\bibinfo{volume}{21}},
  \bibinfo{pages}{612 } (\bibinfo{year}{1976}).

\bibitem[{\citenamefont{Northcliffe and Schilling}(1970)}]{northcliffe}
\bibinfo{author}{\bibfnamefont{L.}~\bibnamefont{Northcliffe}} \bibnamefont{and}
  \bibinfo{author}{\bibfnamefont{R.}~\bibnamefont{Schilling}},
  \bibinfo{journal}{Atomic Data and Nuclear Data Tables}
  \textbf{\bibinfo{volume}{7}}, \bibinfo{pages}{233 } (\bibinfo{year}{1970}),
  ISSN \bibinfo{issn}{0092-640X},
  \urlprefix\url{http://www.sciencedirect.com/science/article/pii/S0092640X7080016X}.

\bibitem[{\citenamefont{Kratz et~al.}(2013)\citenamefont{Kratz, Sch\"adel, and
  G\"aggeler}}]{kratz13}
\bibinfo{author}{\bibfnamefont{J.~V.} \bibnamefont{Kratz}},
  \bibinfo{author}{\bibfnamefont{M.}~\bibnamefont{Sch\"adel}},
  \bibnamefont{and} \bibinfo{author}{\bibfnamefont{H.~W.}
  \bibnamefont{G\"aggeler}}, \bibinfo{journal}{Phys. Rev. C}
  \textbf{\bibinfo{volume}{88}}, \bibinfo{pages}{054615}
  (\bibinfo{year}{2013}),
  \urlprefix\url{http://link.aps.org/doi/10.1103/PhysRevC.88.054615}.

\bibitem[{\citenamefont{Riedel and N\"orenberg}(1979)}]{reidel79}
\bibinfo{author}{\bibfnamefont{C.}~\bibnamefont{Riedel}} \bibnamefont{and}
  \bibinfo{author}{\bibfnamefont{W.}~\bibnamefont{N\"orenberg}},
  \bibinfo{journal}{Zeitschrift f\"ur Physik A Atoms and Nuclei}
  \textbf{\bibinfo{volume}{290}}, \bibinfo{pages}{385} (\bibinfo{year}{1979}),
  ISSN \bibinfo{issn}{0939-7922},
  \urlprefix\url{http://dx.doi.org/10.1007/BF01408400}.

\bibitem[{\citenamefont{Zagrebaev and
  Greiner}(2008{\natexlab{b}})}]{zaggy08prl}
\bibinfo{author}{\bibfnamefont{V.}~\bibnamefont{Zagrebaev}} \bibnamefont{and}
  \bibinfo{author}{\bibfnamefont{W.}~\bibnamefont{Greiner}},
  \bibinfo{journal}{Phys. Rev. Lett.} \textbf{\bibinfo{volume}{101}},
  \bibinfo{pages}{122701} (\bibinfo{year}{2008}{\natexlab{b}}),
  \urlprefix\url{http://link.aps.org/doi/10.1103/PhysRevLett.101.122701}.

\bibitem[{\citenamefont{Kozulin et~al.}(2012)\citenamefont{Kozulin, Vardaci,
  Knyazheva, Bogachev, Dmitriev, Itkis, Itkis, Knyazev, Loktev, Novikov
  et~al.}}]{kozulin12}
\bibinfo{author}{\bibfnamefont{E.~M.} \bibnamefont{Kozulin}},
  \bibinfo{author}{\bibfnamefont{E.}~\bibnamefont{Vardaci}},
  \bibinfo{author}{\bibfnamefont{G.~N.} \bibnamefont{Knyazheva}},
  \bibinfo{author}{\bibfnamefont{A.~A.} \bibnamefont{Bogachev}},
  \bibinfo{author}{\bibfnamefont{S.~N.} \bibnamefont{Dmitriev}},
  \bibinfo{author}{\bibfnamefont{I.~M.} \bibnamefont{Itkis}},
  \bibinfo{author}{\bibfnamefont{M.~G.} \bibnamefont{Itkis}},
  \bibinfo{author}{\bibfnamefont{A.~G.} \bibnamefont{Knyazev}},
  \bibinfo{author}{\bibfnamefont{T.~A.} \bibnamefont{Loktev}},
  \bibinfo{author}{\bibfnamefont{K.~V.} \bibnamefont{Novikov}},
  \bibnamefont{et~al.}, \bibinfo{journal}{Phys. Rev. C}
  \textbf{\bibinfo{volume}{86}}, \bibinfo{pages}{044611}
  (\bibinfo{year}{2012}),
  \urlprefix\url{http://link.aps.org/doi/10.1103/PhysRevC.86.044611}.

\bibitem[{\citenamefont{Savard}()}]{savard14}
\bibinfo{author}{\bibfnamefont{G.}~\bibnamefont{Savard}},
  \bibinfo{note}{\uppercase{ATLAS} Users Meeting, Argonne National Laboratory,
  May 15-16, 2014.}

\bibitem[{\citenamefont{Pollarolo}({\natexlab{a}})}]{pollarolo02}
\bibinfo{author}{\bibfnamefont{G.}~\bibnamefont{Pollarolo}},
  \bibinfo{note}{\uppercase{EURISOL} 2nd Town Meeting at Abamo, 24-25 January,
  2002.}

\bibitem[{\citenamefont{Sch{\"a}del}()}]{schadel10}
\bibinfo{author}{\bibfnamefont{M.}~\bibnamefont{Sch{\"a}del}},
  \bibinfo{note}{\uppercase{IRIS} 10 Workshop, GSI, Darmstadt, Germany, March
  1, 2010.}

\bibitem[{\citenamefont{Pollarolo}({\natexlab{b}})}]{pollarolo}
\bibinfo{author}{\bibfnamefont{G.}~\bibnamefont{Pollarolo}},
  \bibinfo{note}{private communication.}

\bibitem[{\citenamefont{Loveland}()}]{loveland14}
\bibinfo{author}{\bibfnamefont{W.}~\bibnamefont{Loveland}},
  \bibinfo{note}{private communication.}

\bibitem[{\citenamefont{Bass}(1974)}]{bass74}
\bibinfo{author}{\bibfnamefont{R.}~\bibnamefont{Bass}},
  \bibinfo{journal}{Nuclear Physics A} \textbf{\bibinfo{volume}{231}},
  \bibinfo{pages}{45 } (\bibinfo{year}{1974}), ISSN \bibinfo{issn}{0375-9474},
  \urlprefix\url{http://www.sciencedirect.com/science/article/pii/0375947474902929}.

\end{thebibliography}


\begin{thebibliography}{5}
\expandafter\ifx\csname natexlab\endcsname\relax\def\natexlab#1{#1}\fi
\expandafter\ifx\csname bibnamefont\endcsname\relax
  \def\bibnamefont#1{#1}\fi
\expandafter\ifx\csname bibfnamefont\endcsname\relax
  \def\bibfnamefont#1{#1}\fi
\expandafter\ifx\csname citenamefont\endcsname\relax
  \def\citenamefont#1{#1}\fi
\expandafter\ifx\csname url\endcsname\relax
  \def\url#1{\texttt{#1}}\fi
\expandafter\ifx\csname urlprefix\endcsname\relax\def\urlprefix{URL }\fi
\providecommand{\bibinfo}[2]{#2}
\providecommand{\eprint}[2][]{\url{#2}}

\bibitem[{\citenamefont{Yanez and Loveland}(2015)}]{PhysRevC.91.044608}
\bibinfo{author}{\bibfnamefont{R.}~\bibnamefont{Yanez}} \bibnamefont{and}
  \bibinfo{author}{\bibfnamefont{W.}~\bibnamefont{Loveland}},
  \bibinfo{journal}{Phys. Rev. C} \textbf{\bibinfo{volume}{91}},
  \bibinfo{pages}{044608} (\bibinfo{year}{2015}),
  \urlprefix\url{https://link.aps.org/doi/10.1103/PhysRevC.91.044608}.

\bibitem[{\citenamefont{Myers and \ifmmode \acute{S}\else
  \'{S}\fi{}wia\ifmmode~\mbox{\c{}}\else \c{}\fi{}tecki}(1999)}]{myers99}
\bibinfo{author}{\bibfnamefont{W.~D.} \bibnamefont{Myers}} \bibnamefont{and}
  \bibinfo{author}{\bibfnamefont{W.~J.} \bibnamefont{\ifmmode \acute{S}\else
  \'{S}\fi{}wia\ifmmode~\mbox{\c{}}\else \c{}\fi{}tecki}},
  \bibinfo{journal}{Phys. Rev. C} \textbf{\bibinfo{volume}{60}},
  \bibinfo{pages}{014606} (\bibinfo{year}{1999}),
  \urlprefix\url{http://link.aps.org/doi/10.1103/PhysRevC.60.014606}.

\bibitem[{\citenamefont{Ignatyuk et~al.}(1976)\citenamefont{Ignatyuk, Itkis,
  Okolovich, Smirenkin, and Tishin}}]{ignatyuk75}
\bibinfo{author}{\bibfnamefont{A.}~\bibnamefont{Ignatyuk}},
  \bibinfo{author}{\bibfnamefont{M.}~\bibnamefont{Itkis}},
  \bibinfo{author}{\bibfnamefont{V.}~\bibnamefont{Okolovich}},
  \bibinfo{author}{\bibfnamefont{G.}~\bibnamefont{Smirenkin}},
  \bibnamefont{and} \bibinfo{author}{\bibfnamefont{A.}~\bibnamefont{Tishin}},
  \bibinfo{journal}{Sov. J. Nucl. Phys.} \textbf{\bibinfo{volume}{21}},
  \bibinfo{pages}{612 } (\bibinfo{year}{1976}).

\bibitem[{\citenamefont{Lestone}(1995)}]{PhysRevC.52.1118}
\bibinfo{author}{\bibfnamefont{J.~P.} \bibnamefont{Lestone}},
  \bibinfo{journal}{Phys. Rev. C} \textbf{\bibinfo{volume}{52}},
  \bibinfo{pages}{1118} (\bibinfo{year}{1995}),
  \urlprefix\url{https://link.aps.org/doi/10.1103/PhysRevC.52.1118}.

\bibitem[{\citenamefont{Sierk}(1986)}]{sierk86}
\bibinfo{author}{\bibfnamefont{A.~J.} \bibnamefont{Sierk}},
  \bibinfo{journal}{Phys. Rev. C} \textbf{\bibinfo{volume}{33}},
  \bibinfo{pages}{2039} (\bibinfo{year}{1986}),
  \urlprefix\url{http://link.aps.org/doi/10.1103/PhysRevC.33.2039}.

\end{thebibliography}

\newpage
\newpage

\begin{table}[tbp]
\caption{\label{tab1}List of reactions simulated with GRAZING-F. The Coulomb and interaction barriers are calculated within the Bass model \cite{bass74}. Cross sections are taken from simulations. The last column gives the reference to the experimental data if it exists.}
\begin{tabular}{lccccccccc}
\hline
\hline
Reaction & $E_{lab}$ (MeV) & $E_{c.m}$ & $V_{C}$ (MeV) & $V_{int}$ (MeV) & $E_{c.m}/V_C$ &  $\sigma^{transfer}$ (mb) & $\sigma^{transfer}_{fission}$ (mb) & Ref. \\
\hline

$^{136}$Xe+$^{208}$Pb & 701.4  & 423.0 & 423.5 & 430.5 & 1.00 & 2010 & 7   & \cite{kozulin12} \\
                    & 746.3  & 450.0 & 423.5 & 430.5 & 1.06 & 2340 & 29  & \\
                    & 872.8  & 526.0 & 423.5 & 430.5 & 1.24 & 2700 & 122 & \cite{kozulin12} \\
                    & 1024.3 & 617.0 & 423.5 & 430.5 & 1.46 & 2900 & 268 & \cite{kozulin12} \\

$^{136}$Xe+$^{198}$Pt & 1224.0 & 604.9 & 405.9 & 412.1 & 1.49 & 5340 & 29 & \\

$^{86}$Kr+$^{248}$Cm & 435.0 & 323.0 & 340.4 & 344.3 & 0.95 & 6360 & 46   & \cite{moody86} \\
                   & 457.0 & 339.3 & 340.4 & 344.3 & 1.00 & 6590 & 210  & \cite{moody86} \\
                   & 520.0 & 385.1 & 340.4 & 344.3 & 1.13 & 7000 & 3880 & \cite{moody86} \\
                   & 667.4 & 494.0 & 340.4 & 344.3 & 1.45 & 7540 & 6700 & \\

$^{94}$Kr+$^{248}$Cm & 677.8 & 490.0 & 336.9 & 340.7 & 1.45 & 8300 & 7450 & \\

$^{136}$Xe+$^{244}$Pu & 826.0 & 528.8 & 473.9 & 482.7 & 1.12 & 2750 & 1510 & \cite{moody86} \\

$^{129}$Xe+$^{248}$Cm & 780.0 & 511.6 & 486.0 & 495.3 & 1.05 & 7330 & 860 & \cite{welch87} \\

$^{132}$Xe+$^{248}$Cm & 782.0 & 508.9 & 484.6 & 493.8 & 1.05 & 7090 & 670 & \\
                    & 805.0 & 523.8 & 484.6 & 493.8 & 1.08 & 7300 & 800 & \cite{welch87} \\

$^{136}$Xe+$^{248}$Cm & 769.0 & 496.6 & 482.7 & 492.0 & 1.03 & 7330 & 260 & \cite{moody86} \\
                    & 785.0 & 505.6 & 482.7 & 492.0 & 1.05 & 7050 & 810 &  \\

$^{144}$Xe+$^{248}$Cm & 800.0 & 504.7 & 479.2 & 488.3 & 1.05 & 7150 & 160 & \\

$^{136}$Xe+$^{249}$Cf & 749.0 & 483.1 & 492.5 & 502.1 & 0.98 & 2290 & 1680 & \cite{gregorich87} \\
                   & 813.0 & 524.3 & 492.5 & 502.1 & 1.06 & 2710 & 1430 & \cite{gregorich87} \\
                   & 877.0 & 565.4 & 492.5 & 502.1 & 1.15 & 2910 & 1670 & \cite{gregorich87} \\

$^{238}$U+$^{238}$U & 2059.0 & 1024.8 & 735.2 & 754.6 & 1.39 & 9310 & 4710 & \cite{kratz13} \\
                  & 1785.0 & 899.9 & 735.2 & 754.6 & 1.21 & 9060 & 1760 & \cite{kratz13} \\
                  & 1628.0 & 811.0 & 735.2 & 754.6 & 1.10 & 8930 & 750 & \cite{kratz13} \\
                  & 1545.0 & 769.8 & 735.2 & 754.6 & 1.05 & 8750 & 350 & \cite{kratz13} \\

$^{238}$U+$^{248}$Cm & 1760.0 & 894.6 & 762.6 & 783.2 & 1.17 & 8860 & 1400 & \cite{kratz13} \\

$^{238}$U+$^{249}$Bk & 1587.9 & 809.0 & 770.1 & 791.0 & 1.05 & 8780 & 6500 & \\
                   & 2195.4 & 1117.0 & 770.1 & 791.0 & 1.45 & 9980 & 7960 & \\
\hline
\hline
\end{tabular}
\end{table}

\begin{table}[tbp]
\caption{\label{tab2}Predicted cross sections of unknown U and Np isotopes in the $^{238}$U+$^{238}$U reaction at $E_{lab}=2059$ MeV. Only cross sections $> 100$ pb are listed.}
\begin{tabular}{lccc}
\hline
\hline
Isotope & $\sigma$ ($\mu$b) \\
\hline
$^{243}$U & $21.8$ \\
$^{244}$U & $50.8$ \\
$^{245}$U & $3.7$  \\
$^{246}$U & $2.4$  \\
$^{247}$U & $0.25$ \\
$^{245}$Np & $2.9$ \\
$^{246}$Np & $0.22$ \\
$^{247}$Np & $0.35$ \\
\hline
\hline
\end{tabular}
\end{table}

\newpage

\begin{table}[tbp]
\caption{\label{tab3}Maximum production rates of $N=126$ isotopes in the $^{136}$Xe+$^{198}$Pt reaction at $E_{lab}=9$ MeV/A simulated with GRAZING-F assuming a beam current of 1 p$\mu$A and a target thickness equivalent to the range from the entrance energy to the interaction barrier.}
\begin{tabular}{lc}
\hline
\hline
Isotope & $R_{max}$ (s$^{-1}$) \\ \hline
$^{204}_{78}$Pt & $2.6 \times 10^{6}$ \\
$^{203}_{77}$Ir & $4.7 \times 10^{5}$\\
$^{202}_{76}$Os & $5.5 \times 10^{4}$\\
$^{201}_{75}$Re & $4.0 \times 10^{3}$ \\ 
\hline
\hline
\end{tabular}
\end{table}

\newpage

\begin{table}[tbp]
\caption{\label{tab5}Yields $N$ of U and Pu isotopes in the $^{238}$U+$^{249}$Bk reaction at $E_{c.m.}=809$ MeV simulated with GRAZING-F assuming a beam current of 100 pnA, a target thickness of $0.3$ mg/cm$^2$ and 1 day irradiation.}
\begin{tabular}{lccc}
\hline
\hline
Isotope & $N$ & Isotope & $N$ \\
\hline
$^{244}$U & $1.9\times10^6$ & $^{248}$Pu & $1.6\times10^7$\\
$^{245}$U & $5.7\times10^5$ & $^{249}$Pu & $2.8\times10^6$ \\
$^{246}$U & $1.3\times10^6$ & $^{250}$Pu & $3.2\times10^6$\\
$^{247}$U & $2.9\times10^5$ & $^{251}$Pu & $2.0\times10^5$\\
$^{248}$U & $2.8\times10^5$ & $^{252}$Pu & $3.0\times10^5$\\
$^{249}$U & $6.7\times10^4$ & $^{253}$Pu & $7.0\times10^3$\\
$^{250}$U & $2.5\times10^4$ & $^{254}$Pu & $1.4\times10^4$\\
\hline
\hline
\end{tabular}
\end{table}

\newpage

\end{document}